\documentclass{jfm}

\usepackage{graphicx,epsfig,subfig}
\usepackage{epstopdf}
\usepackage{natbib}
\usepackage{amsmath}
\usepackage{mathtools}
\usepackage{amssymb}
\usepackage{color}
\usepackage{setspace}
\usepackage{mathrsfs}
\usepackage[usenames,dvipsnames]{xcolor}
\usepackage{pstool}
\usepackage[linesnumbered,ruled]{algorithm2e}

\newcommand\zi{\mathrm{i}}

\def\bdot{\raise.2em\hbox to .15em{.}}

\definecolor{gray}{gray}{0.8}
\definecolor{darkgray}{gray}{0.5}

\def\bdotblack{\raise.25em\hbox to .15em{.}}

\title[Periodic orbits and DMD]{Searching turbulence for periodic orbits with dynamic mode decomposition}

\author[J. Page \& R. R. Kerswell]%
{Jacob Page \& Rich R. Kerswell} 

\affiliation{DAMTP, Center for Mathematical Sciences, University of Cambridge, Cambridge, CB3 0WA, UK}

\pubyear{2018}
\volume{}
\pagerange{}
\date{\today}
\begin{document}

\maketitle
\begin{abstract}
    We present a new method for generating robust guesses for unstable periodic orbits (UPOs) by post-processing turbulent
    data using dynamic mode decomposition (DMD).
    The approach relies on the identification of near-neutral, repeated harmonics in the DMD eigenvalue spectrum from 
    which both an estimate for the period of a nearby UPO and a guess for the velocity field can be constructed.
    In this way, the signature of a UPO can be identified in a short time series without the need for a near
    recurrence to occur,
    which is a considerable drawback to recurrent flow analysis, the current state-of-the-art.
    We first demonstrate the method by applying it to a known (simple) UPO and find that the period can be reliably extracted
    even for time windows of length one quarter of the full period.
    We then turn to a long turbulent trajectory, sliding an observation window through the time series and 
    performing many DMD computations. 
    Our approach yields many more converged periodic orbits (including multiple new solutions) than a standard recurrent flow analysis
    of the same data.
    Furthermore, it also yields converged UPOs at points where the recurrent flow analysis flagged a near recurrence but the Newton
    solver did not converge, suggesting that the new approach can be used alongside the old to generate improved initial guesses.
    Finally, we discuss some heuristics on what constitutes a ``good'' time window for the DMD to identify a UPO. 
\end{abstract}

\section{Introduction}
%
Since the discovery of the first unstable period orbit (UPO) buried in a turbulent attractor (plane Couette flow) by
\citet{Kawahara2001}, there has been a surge in interest and a large number of other periodic solutions
found both in this configuration \citep[e.g.][]{Viswanath2007,Cvitanovic2010} and in other canonical turbulent flows
\citep{Chandler2013,Willis2013,Lucas2017}.
The discovery of a large number of UPOs supports a perspective of turbulence
in which the flow is viewed as a trajectory in a very high-dimensional dynamical system, wandering
between unstable exact coherent structures \citep{Kerswell2005,Kawahara2012}.
Individual UPOs 
can offer a good deal of insight into the physical processes sustaining the turbulent flow owing 
to their simple time dependence \citep{Waleffe1997,Wang2007,Hall2010}.
Furthermore, periodic orbit theory \citep{ChaosBook} suggests that statistics of the turbulence can
be predicted from the UPOs if enough of them are found \citep[see the attempts in][]{Chandler2013, Lucas2015}.
Despite this increasing interest, the methods for finding UPOs are somewhat crude and have not changed significantly in the two decades since the first
turbulent solutions were discovered. 

The standard method for finding and converging UPOs begins with a search for near recurrences in data from a numerical simulation.
In such a `recurrent flow analysis', likely UPOs are flagged when the distance in state space (measured with an $L_2$ norm) between
the present and past states drops below a threshold value \citep{Kawahara2001,Viswanath2007,Cvitanovic2010,Chandler2013}.
The resulting set of candidate orbits, each augmented with a guessed period from the time between two similar states, is then 
input into a Newton solver.
The main downside of the approach is that it requires the turbulence to shadow a periodic orbit for at least one full cycle,
and hence it can be increasingly ineffective as the Reynolds number is increased \citep{Chandler2013}.
In addition, the sensitivity of the Newton solver to initial conditions can result in failures at the convergence stage 
when there may in fact be a UPO nearby.
In this paper we introduce a new method based on dynamic mode decomposition (DMD) that goes some way to addressing these issues.

The DMD algorithm was originally invented by \citet{Schmid2010} and can serve as an excellent data-driven method for 
performing global stability analyses in complex geometries, since it requires only raw data in the form of snapshot pairs.
The output of the algorithm is a linear operator that (in a least squares sense over the input data) provides the best mapping between
snapshots separated by a fixed time $\delta t$.
An eigendecomposition of this operator yields a set of `dynamic modes' which evolve exponentially in time, the time dependence being set
by the associated eigenvalues.
Crucially, this approach can identify frequencies of oscillation with corresponding periods that can be far longer than the time window over
which the observations were recorded \citep{Schmid2010,Rowley2009,Schmid2011}.
Since its invention, a number of variants of the algorithm have been proposed and DMD has found applications in areas of science and engineering
beyond fluid mechanics \citep{Jovanovic2014,Williams2015,DMDkutz}.

Despite its apparent linearity, DMD can also be a useful tool in strongly nonlinear flows.
Our interest in applying DMD to turbulent data as a means to identify UPOs rests on the equivalence (under strict conditions)
of the algorithm to the Koopman decomposition \citep{Rowley2009,Tu2014,Williams2015} of a nonlinear dynamical system.
The Koopman operator is an infinite-dimensional, linear operator that propagates observables
of the state forward in time along trajectories of the dynamical system \citep{Koopman1931,Mezic2005},
and a Koopman decomposition of the state leads to a representation in which the nonlinear dynamics are expressed as linear 
superposition of Koopman modes (e.g. fixed velocity fields) which evolve exponentially in time.
Koopman decompositions have been performed analytically for some simple nonlinear ODEs \citep{Bagheri2013,Brunton2016,Rowley2017}
and PDEs \citep{Page2018}.

Applying a Koopman decomposition to a turbulent flow yields a representation
of the state as a superposition of a set of harmonic averages and a broadband continuous spectrum
\citep{Mezic2004,Mezic2005,Mezic2013,Arbabi2017},
but individual simple invariant sets also possess their own \emph{local} Koopman decompositions.
For example, \citet{Mezic2017} has shown that the Koopman eigenvalues for a nonlinear system collapsing onto a limit cycle consists of 
a set of repeated neutral harmonics (the limit cycle's fundamental frequency and higher harmonics) and an infinite lattice 
of decaying eigenvalues which can be determined from linear combinations of the cycle's Floquet multipliers.
However, in systems with multiple exact coherent structures there are multiple local Koopman decompositions around each structure.
These linear representations break down (are no longer convergent) at certain \emph{crossover points} in state space \citep{Page2019}, a fact which impacts 
the ability of DMD to extract the Koopman modes associated with the expansions.
That DMD no longer coincides with Koopman when the observation window includes the crossover point is likely the reason
behind the poor performance of the algorithm when \citet{Bagheri2013} applied it to the `spin up' problem of flow past 
a cylinder.
\citet{Page2019} demonstrated the phenomenon explicitly in a Stuart-Landau equation and numerically along heteroclinic connections in the 
Navier-Stokes equations.
In doing this, it was observed that DMD can extract some of the Koopman eigenvalues associated with the local Koopman decomposition 
around a periodic orbit -- though only if the computation is performed within the `Koopman expansion zone' where the 
local decomposition is convergent.
Here, we go further and layout how DMD can be applied to turbulent data to identify and converge UPOs.


The remainder of this paper is organised as follows.
In \S2 we outline a procedure for performing and post-processing DMD calculations to generate guesses for periodic orbits
and demonstrate its utility by applying it to a gently periodic edge state.
In \S3 the method is then used to search for periodic orbits close to a long turbulent 
trajectory and is compared to recurrent flow analysis.
We examine some of the DMD guesses further in \S4, and identify characteristics of ``good'' initial guesses.
Finally, concluding remarks are provided in \S5. 

\section{Methodology}
In this paper we present a new method for identifying periodic orbits using DMD applied to short turbulent trajectories
and contrast it to recurrent flow analysis, the current state-of-the-art. 
To do this we post-process numerical simulation data using both methods and attempt to converge the resulting UPO guesses 
via Newton-Krylov iteration. 
Throughout, the flow configuration is plane Couette flow in a minimal flow unit \citep{Hamilton1995} non-dimensionalized by the upper/lower
plate speed $U_0$ and the channel half-height $d$.
The Reynolds number is fixed at $Re:=U_0 d/\nu=400$, while the
exact horizontal dimensions of the periodic computational domain are $(L_x,L_y) = (1.755\pi, 1.2\pi)$ to match \citet{Kawahara2001}.

In our direct numerical simulations, the Navier-Stokes equations are solved using a fractional-step method with an implicit Crank-Nicholson 
scheme for the diffusive terms and a third-order Runge-Kutta scheme for the advection terms \citep{Dubief2005}.
Second-order finite differences are used in all three spatial directions with grid stretching applied in the vertical. 
For the majority of results the grid resolution is $N_x\times N_y\times N_z = 96 \times 96 \times 129$. 
Note that the slight difference in box size from the $L_x=1.75\pi$ box used by \citet{Cvitanovic2010} 
(and the solutions catalogued at \texttt{channelflow.org})
precludes a direct comparison with this extensive set of UPOs.
Altering our box size to $1.75\pi$ indicates that some of UPOs we find are known solutions listed at \texttt{channelflow.org}.
At our resolution we can match the periods of these known periodic orbits to within a relative error of $\sim 0.1$\%, 
and have verified that this error can be reduced by
further increasing our resolution. We make every effort to connect our results to previously discovered exact solutions (see table \ref{tab:DMD}).

\begin{figure}
    \centering
    \includegraphics[width=0.42\textwidth]{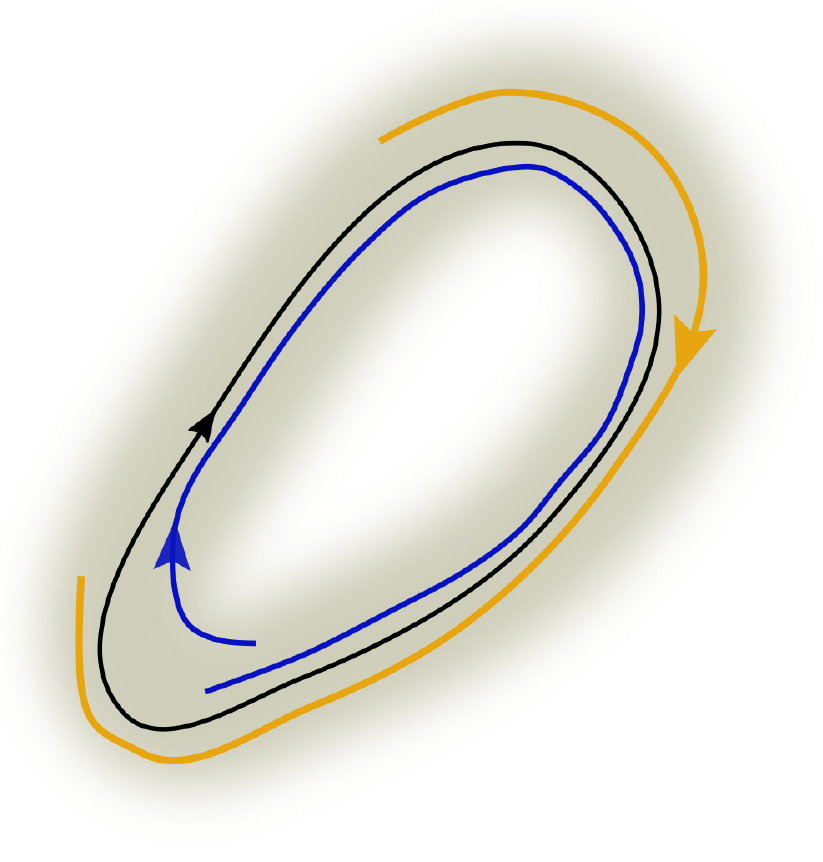}
    \vspace{-5mm}
    \caption{Cartoon of a periodic orbit (black) and two trajectories where DMD provides a useful alternative to recurrent flow analysis.
    The orange line is a segment of an orbit which does not shadow the UPO for a full period and hence is useless for recurrent flow analysis.
    The blue line does shadow the UPO for a full cycle, but using the local mimimum in $R(t,T)$ may yield an inaccurate estimate of the period.
    The shaded region identifies the ``Koopman expansion zone'' around the UPO, discussed in the text.}
    \label{fig:schematic}
\end{figure}
In a recurrent flow analysis \citep[e.g.][]{Chandler2013}, guesses for periodic orbits are identified via local minima in an $L_2$ norm,
\begin{equation}
    R(t,T) := \frac{\|\mathbf u(t+T) - \mathbf u(t)\|}{\|\mathbf u(t)\|},
    \label{eqn:recur}
\end{equation}
which fall below some threshold $R_{\text{thresh}}$. 
If the minima occur at times $t_i$, then the guesses for periodic orbits are simply the corresponding states $\{\mathbf u(t_i)\}$,
each augmented with a guessed period $T_g=T_i$ -- the future time at which the local minimum in $R(t,T)$ occurs.
Note that in the form (\ref{eqn:recur}) we are comparing \emph{future} states at $t+T$ with present states at $t$ in order to allow for
comparison to results from DMD with observation windows starting at time $t$. 

As discussed in the introduction, there are shortcomings to this approach for generating guesses for periodic orbits.
For instance, given a near recurrence, the (future) time $T$ at which the minimum value of $R(t,T)$ occurs may differ significantly 
from the period of the nearby UPO. 
It is plausible that the sensitivity of Newton-Raphson to 
initial conditions could result in a lack of convergence even if there is indeed a structure nearby. 
In these situations, DMD may be a useful tool since it can be used to extract a set of modes associated with `background' periodic motion, resulting in a 
more robust initial condition for the root-finder and a more informed estimate of the period.
Perhaps more importantly, DMD does not require a near recurrence to occur before it can `sense' periodic motion since the algorithm allows for the extraction 
of frequencies corresponding to oscillations far longer than the length of observation window \citep{Schmid2010}.
Both of these scenarios are sketched in figure \ref{fig:schematic}.

In DMD, vector \emph{observables} of the state at discrete times $\{t_j\}$ are stored in a data matrix \citep{Schmid2010, Tu2014}, 
\begin{equation}
    \boldsymbol \Psi^t = \begin{bmatrix} \boldsymbol \psi(\mathbf u(t_1)) & \boldsymbol \psi(\mathbf u(t_2)) & \cdots & 
    \boldsymbol \psi(\mathbf u(t_M))\end{bmatrix},
\end{equation}
where the observable functionals that make up the elements of $\boldsymbol \psi$ are a design choice, often motivated by knowledge of the underlying dynamical 
system \citep[e.g.][]{Williams2015}. 
A second data matrix is formed with observations made a time $\delta t$ (another design choice) later, 
\begin{equation}
    \boldsymbol \Psi^{t+\delta t} = \begin{bmatrix} \boldsymbol \psi(\mathbf u(t_1+\delta t)) & \boldsymbol \psi(\mathbf u(t_2+\delta t)) & \cdots & 
    \boldsymbol \psi(\mathbf u(t_M+\delta t))\end{bmatrix},
\end{equation}
where we refer to corresponding columns in $\boldsymbol \Psi^t$ and $\boldsymbol \Psi^{t+\delta t}$ as a `snapshot pair'.
The DMD operator is then the best-fit (in a least-squares sense) linear operator that maps between the data matrices, 
\begin{equation}
    \hat{\mathbf K} := \boldsymbol \Psi^{t+\delta t}\left(\boldsymbol \Psi^{t}\right)^+
\end{equation}
where the superscript $+$ identifies the Moore-Penrose pseudo-invserve, 
accomplished via an SVD of the data matrix $\boldsymbol \Psi^{t} = \mathbf U\boldsymbol \Sigma\mathbf W^H$.
In reality the dimensionality of the problem requires that $\hat{\mathbf K}$ is never computed directly,
but instead the $r\times r$ (where $r$ is the rank of $\boldsymbol \Psi^{t}$) matrix $\mathbf U^H\hat{\mathbf K}\mathbf U$ is 
used -- the projection of the DMD matrix onto POD (Proper Orthogonal Decomposition) modes.

To search for periodic orbits we will be applying DMD to relatively short turbulent trajectories ($O(100)$ advective time units with
$M\sim 100$ snapshot pairs), and hence the data matrix will likely be full rank $r=M$.  
This tends to lead to overfitting and unphysical DMD modes, a behaviour which can be avoided through a variety of techniques, 
e.g. by either performing a low-rank truncation or by introducing a sparsity constraint \citep{Jovanovic2014}.
Rather than selecting a threshold singular value or modifying the DMD algorithm, we instead attempt to find a DMD operator 
that can map between velocity fluctuations above a certain threshold by adding a small amount of white noise to our observable vector,
\begin{equation}
    \boldsymbol \psi(\mathbf u) := \mathbf u - \mathbf u_C + \boldsymbol{\delta w}_{\epsilon}.
\end{equation}
where $\mathbf u_C:=z\mathbf e_x$ is the laminar base state and $\boldsymbol{\delta w}_{\epsilon}$ is the white noise of amplitude $\epsilon$.
The minimum singular value $\sigma_{min}$ scales proportionally with $\epsilon$.
Throughout, we set $\epsilon = 10^{-7}$, though we have verified that our results are robust under modest changes in $\epsilon$ (plus or minus a decade). 

Once the (projected) DMD operator is computed, an eigendecomposition can be performed.
Under certain conditions \citep[see][]{Williams2015,Rowley2017,Page2019} the output of DMD coincides with a Koopman analysis.
In these cases, approximate Koopman modes for the observable $\boldsymbol \psi(\mathbf u)$ can be obtained as right eigenvectors of $\hat{\mathbf K}$
(DMD modes),
\begin{equation}
    \hat{\mathbf K}\mathbf v_j = e^{\lambda_j \delta t} \mathbf v_j.
\end{equation}
Note that if $\mathbf v'$ is a right eigenvector of $\mathbf U^H\hat{\mathbf K}\mathbf U$, the DMD mode is obtained from 
$\mathbf v = \mathbf U \mathbf v'$ \citep{Schmid2010}. 
Koopman eigenfunctions are related to the left eigenvectors of $\hat{\mathbf K}$, $\{\mathbf w_j\}$, via
\begin{equation}
    \varphi_{\lambda_j}(\mathbf u) = \mathbf w_j^H \boldsymbol \psi(\mathbf u).
    \label{eqn:efns}
\end{equation}


\subsection{Spotting a periodic orbit}
\label{sub:upo_id}
In a dynamical system with a single attracting periodic orbit, \citet{Mezic2017} has shown that the spectrum of the associated Koopman operator consists 
of a set of repeated harmonics (multiples of the fundamental frequency of the limit cycle) along with an infinite lattice of decaying eigenvalues
that can be computed from the Floquet multipliers.
In the turbulent Couette flow considered in this work, all periodic orbits are unstable and each orbit will have an associated locally convergent
Koopman decomposition \citep{Page2019} of a similar form to that derived by \citet{Mezic2017} but with unstable eigenvalues related to the set of 
unstable Floquet multipliers.
While DMD on a trajectory shadowing the UPO for (potentially less than) one cycle is unlikely to be able to identify the stability properties 
of the nearby structure, the results presented in \citet{Page2019} indicate that the signature of the periodic orbit -- its neutral harmonics -- in the DMD 
results can remain even for relatively short time windows.

In previous studies involving DMD on flows collapsing onto limit cycles \citep[e.g.][]{Bagheri2013, Page2019}, identifying the underlying period in the 
DMD eigenvalue spectrum is straightfoward either because the period is known beforehand or the DMD can be run over an indefinitely long time horizon
due to the stability of the exact coherent structure.  
However, in a turbulent flow it is unknown a-priori at any given time whether we are shadowing a periodic orbit, and if so what its period is.
Therefore, given an eigenvalue spectrum from a DMD calculation on a segment of a turbulent orbit we would like to both 
(i) estimate the period of a nearby UPO (if one exists) and (ii) assess the quality of our ``guess'' to make an informed decision about 
whether there is indeed a nearby structure.

For step (i), the estimation of a UPO period, we proceed as follows:
A maximum cut-off growth/decay rate, $\lambda_r^{\text{max}}$, is chosen to select a subset of the DMD modes, $\tilde{\Lambda}:=\{\lambda \; | \; |Re(\lambda)|\leq \lambda_r^{\text{max}}\}$.
The truncated set of eigenvalues $\tilde{\Lambda}$ contains at least one mode with $Im(\lambda) = 0$ and pairs of complex conjugate eigenvalues.
Of the modes with $Im(\lambda)=0$, the one with the smallest real component is assumed to be the time average of a nearby UPO,
while the first $n$ modes with positive imaginary components are assumed to constitute the UPO's fundamental frequency and its $n-1$ higher harmonics.
Under this assumption, a fundamental frequency can be estimated via
\begin{equation}
    \omega_f(n) := \frac{2}{n(n+1)}\sum_{j=1}^n\omega_j,
    \label{eqn:freq_estimate}
\end{equation}
where $\omega_j := Im(\lambda_j)$.
The guess for a velocity field at a point on the UPO is then built from the time average mode and the first $n$ complex conjugate pairs (this procedure is described in more detail in \S\ref{sec:build} below), while the period is esimated as $T_g = 2\pi/\omega_f$. 
For step (ii), the estimate of how good such a guess is expected to be, we determine the degree to which the finite-frequency modes approximate a series of harmonics, 
\begin{equation}
    \varepsilon_{\omega}(n) := \frac{1}{n|\omega_f|^2}\sum_{j=1}^n |\omega_j - j\omega_f|^2,
    \label{eqn:spec_err}
\end{equation}
with $\varepsilon_{\omega}=0$ for a perfectly periodic signal.
In the following analysis, DMD results for which values of $\varepsilon_{\omega}$ fall below a prescribed threshold are taken to indicate the likely existence
of a nearby UPO.

\subsection{Motivating example: Gently periodic edge state}
\label{sec:edge}
\begin{figure}
    \centering
    \includegraphics[width=0.47\textwidth]{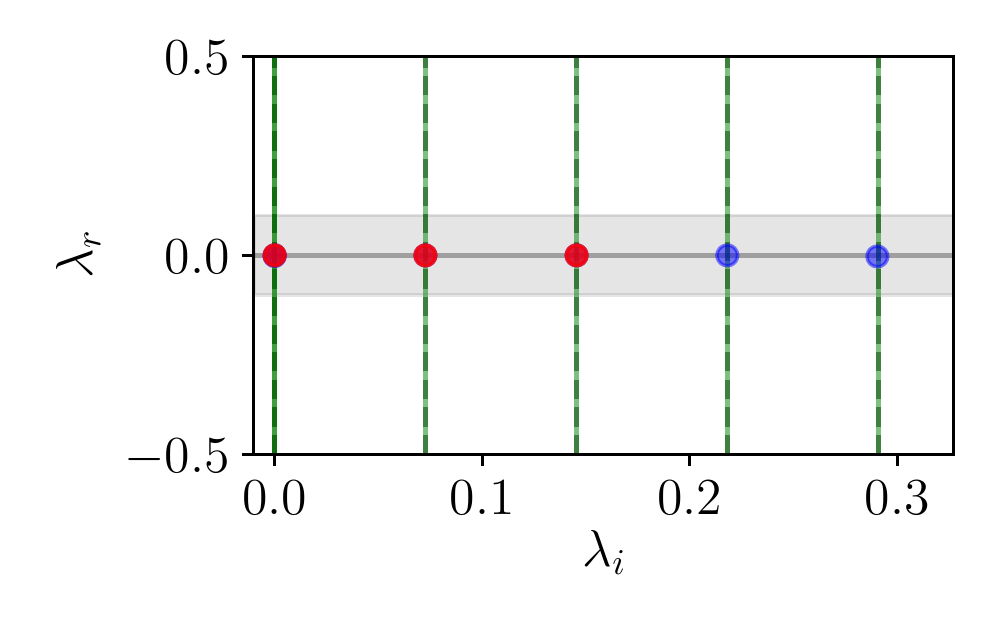}
    \includegraphics[width=0.47\textwidth]{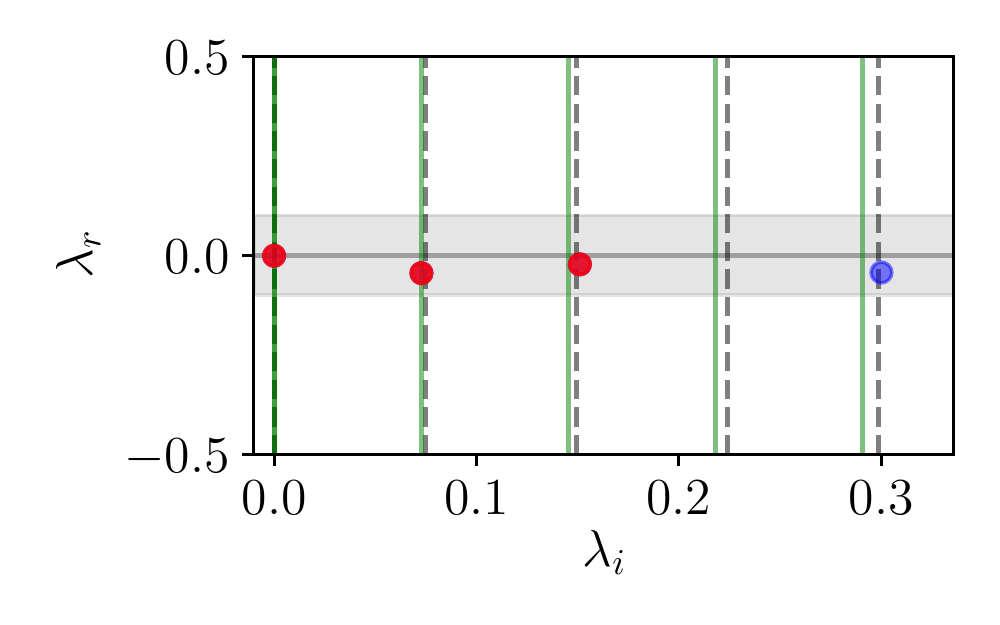}
    \caption{DMD eigenvalues obtained from observations on the ``edge trajectory'' described in the text. 
    (Left) Observation window length matches the period of the edge UPO, $T_w=T_{\omega}$. 
    (Right) Observation window is a quarter of the UPO period, $T_w=T_{\omega}/4$.
    Green lines identify integer multiples of the fundamental frequency of the nearby UPO, $2\pi/T_{\omega}$.
    Dashed black lines correspond to the fundamental frequency estimated from the first two harmonics (highlighted red) in the DMD
    via equation (\ref{eqn:freq_estimate}).
    The shaded region indicates a threshold $\lambda_r^{\text{max}} = 0.1$.
    In both calculations $\delta t = 1$, $M=50$.}
    \label{fig:edge_evs}
\end{figure}
\begin{figure}
    \centering
    \includegraphics[width=0.529\textwidth]{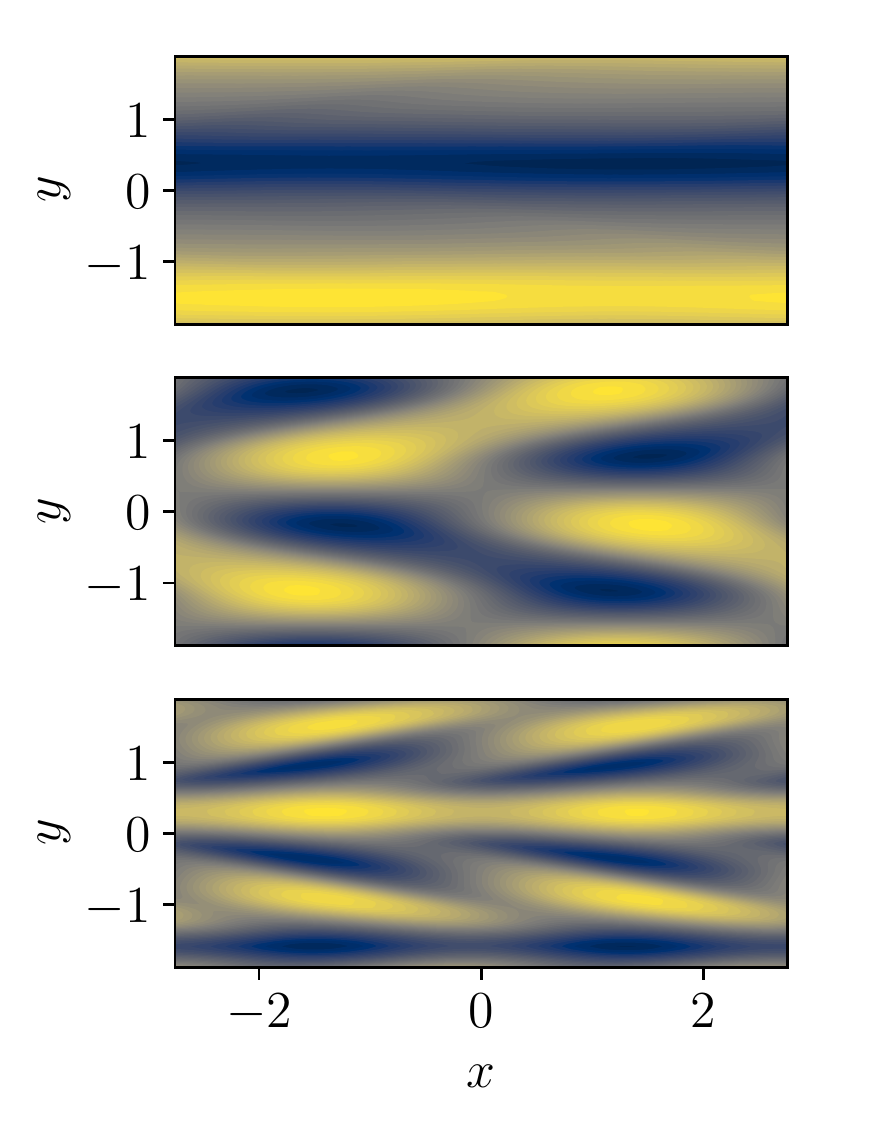}
    \includegraphics[width=0.431\textwidth]{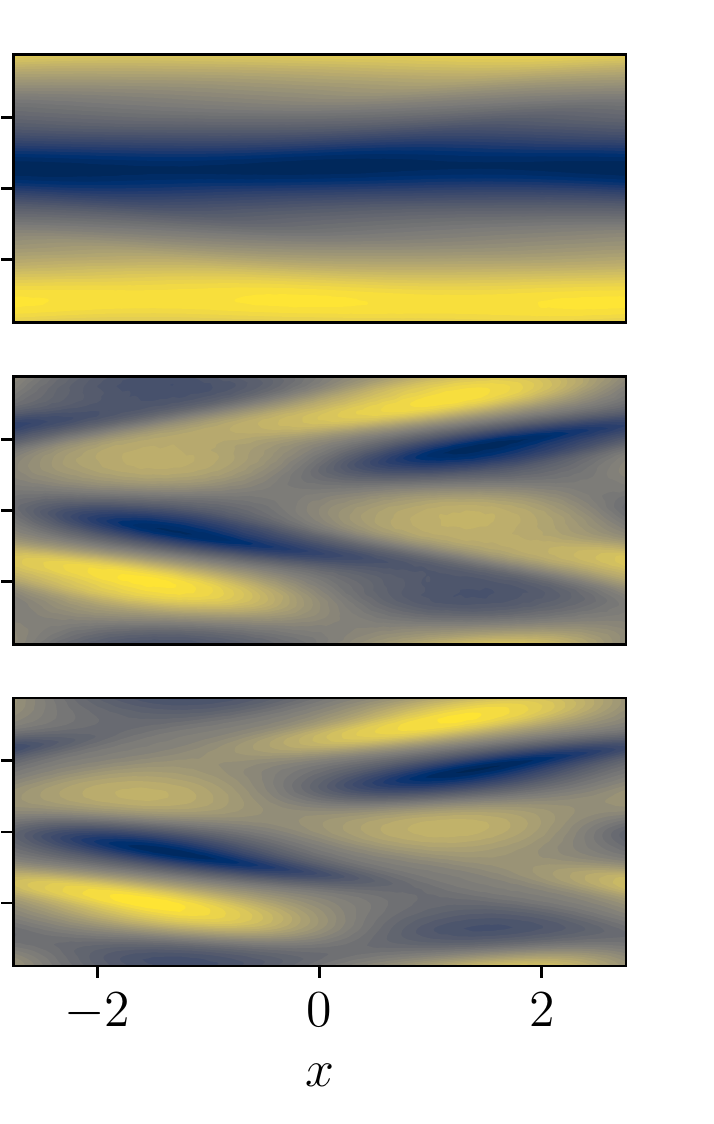}
    \caption{Streamwise velocity on a plane midway between the two walls for the DMD modes corresponding to the eigenvalue spectra reported in figure \ref{fig:edge_evs}. The neutral mode and the first and second harmonics are shown. (Left) $T_w=T_{\omega}$, (right) $T_w=T_{\omega}/4$.}
    \label{fig:edge_streaks}
\end{figure}
To demonstrate the mechanics of the approach outlined above we first apply it to a trajectory that shadows the simplest UPO in this box (the edge state) for 
several complete cycles before becoming turbulent.
This `gentle' periodic orbit was first discovered by \citet{Kawahara2001} and has a period $T_{\omega} \approx 86.4$ in this box 
\citep[see also][]{Kawahara2005}. 
The trajectory is obtained via repeated bisection between relaminarising and turbulent trajectories \citep{Schneider2007} and we store
one full period at a resolution of $\Delta t=0.1$.

DMD spectra from two calculations on this dataset are reported in figure \ref{fig:edge_evs}.
The two calculations are identical in all respects apart from the length of the observation window, $T_w$, over which the DMD is performed.
In the first case, the length of the observation window matches the period of the underlying UPO, $T_w\approx T_{\omega}$, and
the DMD spectrum consists of a single neutral mode and a series of near-perfect harmonics.
The fundamental frequency estimated from the first purely imaginary mode and the first harmonic matches the fundamental frequency of the UPO,
while the associated value $\varepsilon_{\omega}$ is very close to zero.
This result is entirely expected owing to the observation of a full period and the proximity of the trajectory to the UPO. 

More interestingly, the results of the DMD calculation on the window $T_w=T_{\omega}/4$
also clearly show the signature of the UPO despite the short observation time.
The eigenvalues are now slightly off the neutral line $\lambda_r=0$, but there are still clearly repeated harmonics (the third harmonic $\sim 3\omega_f$ is missing).
The period estimated from the first two harmonics is very close to the true period of the UPO, and the corresponding value of $\varepsilon_{\omega} \lesssim 10^{-3}$. 

The two DMD calculations are compared further in figure \ref{fig:edge_streaks}.
The neutral DMD mode obtained over the short time window $T_{\omega}/4$ is qualitatively similar to the true neutral Koopman mode associated with the orbit 
(the DMD mode from the observation window $T_w = T_{\omega}$), 
though the performance is notably worse for the higher harmonics.
However, initial conditions built from these modes (via the procedure described in the following section) still rapidly converge to the UPO
when used as an initial guess in Newton-Krylov routine.

\begin{figure}
    \centering
    \includegraphics[width=0.47\textwidth]{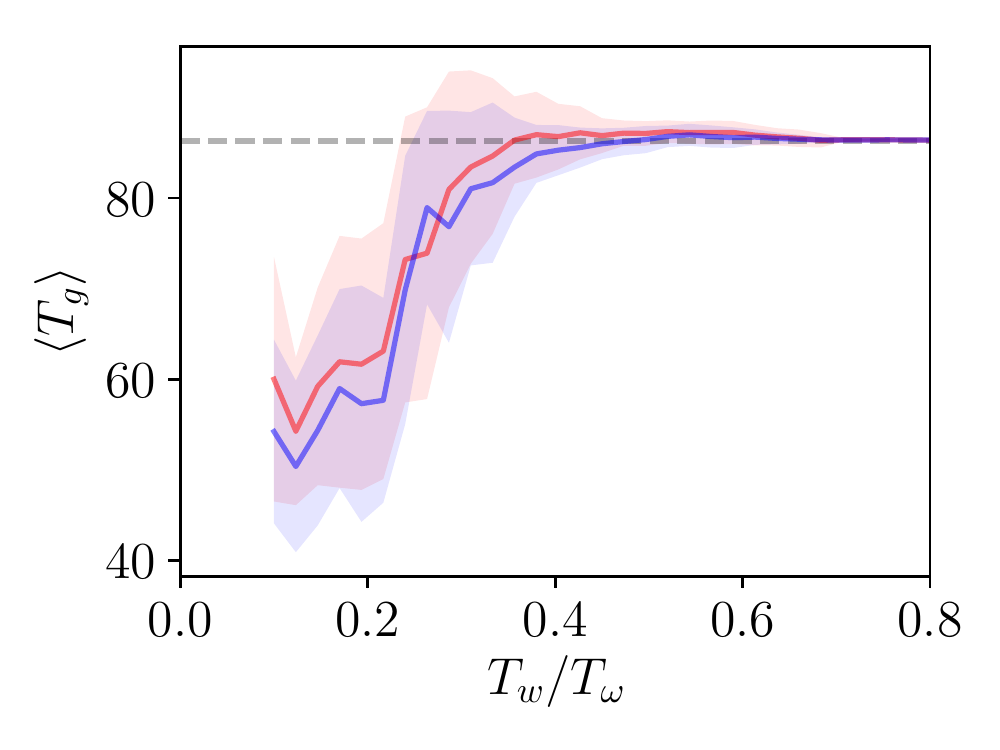}
    \includegraphics[width=0.47\textwidth]{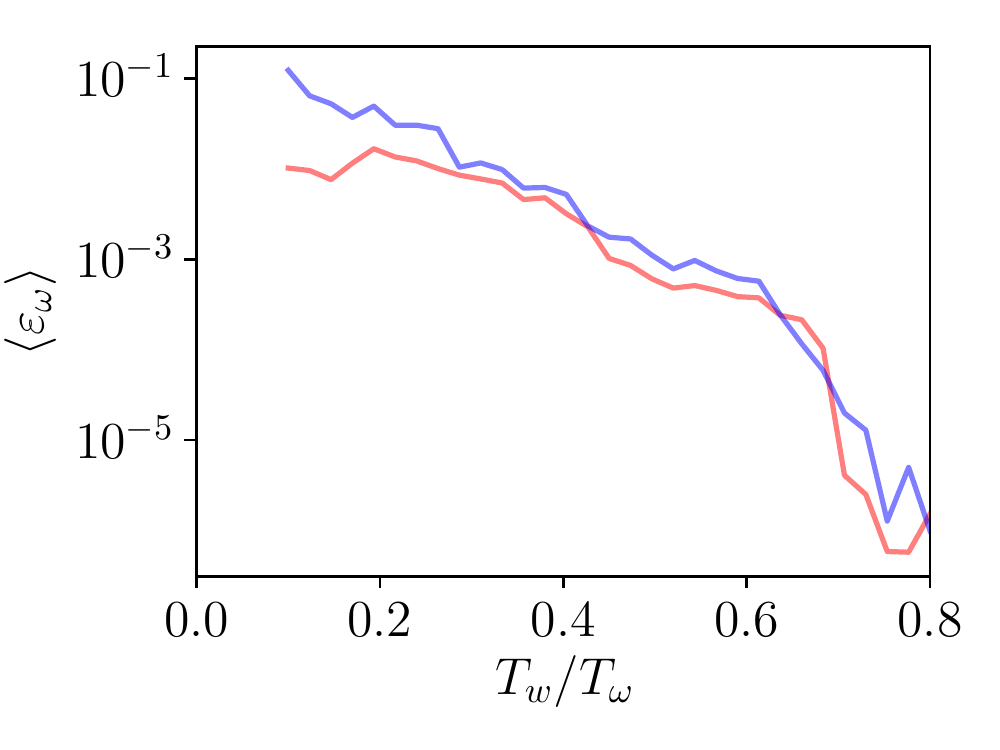}
    \caption{(Left) Average predicted period $T_g=2\pi/\omega_f$ of the edge state UPO as a function of observation window length using estimates 
    based on two (red) and three (blue) harmonics. For each value of $T_w$, averaging is performed over sets of continuous time windows 
    with start times separated by $\Delta t_s=5$. Shaded regions identify $\pm 1$ sample standard deviation. 
    The horizontal dashed line identifies the true period $T_{\omega} = 86.4$. 
    (Right) ``Quality'' of the 
    guessed period as determined by equation (\ref{eqn:freq_estimate}).}
    \label{fig:edge_Tg}
\end{figure}
The performance of shortened observation windows on the edge UPO 
is examined for a wide range of window lengths, each averaged over a large range of possible start times 
$t_s \in [0, T_{\omega})$, in figure \ref{fig:edge_Tg}.
The results indicate that short windows $T_w \lesssim 0.4 T_{\omega}$ can perform well \emph{sometimes}.
On average, the DMD gets progressively worse at identifying the underlying UPO as the length of the observation window is reduced, 
though the accuracy of an individual computation depends subtly on the start time of observations $t_s$ 
as this controls what segment of the UPO is seen by the DMD. 
Importantly, the good results shown for the $T_w=T_{\omega}/4$ window in figure \ref{fig:edge_evs} would not have been obtained if an alternate start time 
had been selected.
General rules of thumb as to what exactly the DMD needs to see in order to provide a robust estimate of the period $T_{\omega}$ of a UPO 
are discussed in \S 4 following the presentation of results from a turbulent computation.

\subsection{Building a guess for a periodic orbit}
\label{sec:build}
While the above results were obtained by examining trajectories that shadow a known orbit, the periodic orbits close to a turbulent trajectory 
will not be known beforehand, and we would like to also use the output of DMD
to build initial `guesses' to supply to a Newton solver if the value of $\varepsilon_{\omega}$ drops below the prescribed threshold.
The approach we adopt is outlined here.

For DMD results where $\varepsilon_{\omega}<\varepsilon_{\omega}^{\text{thresh}}$ for a given number of harmonics $n$, 
the period of the UPO is estimated from the fundamental frequency $\omega_f(n)$, defined in equation (\ref{eqn:freq_estimate}), 
by $T_g = 2\pi/\omega_f$.
The neutral mode and the first $n$ complex conjugate pairs used to estimate $\omega_f(n)$ are then used to construct a guess for the velocity field 
on the UPO,
\begin{equation}
    \mathbf u_g(t) = z\mathbf e_x +  \sum_{j=-n}^n a_j \mathbf v_j e^{\lambda_j t}.
    \label{eqn:u_guess}
\end{equation}
There are a variety of ways to define the unknown coefficients $\{a_j\}$.
We set them such that the average (squared) deviation of the guess from the turbulent trajectory used in the DMD is smallest, 
i.e. by minimising
\begin{equation}
    J(\mathbf a) := \frac{1}{M} \sum_{m=0}^{M-1} \big|\boldsymbol \psi(\mathbf u(\mathbf x,m\delta t)) - \sum_{j=-n}^n a_j\mathbf v_j e^{m\lambda_j\delta t}\big|^2.
\end{equation}
The minimising $\hat{\mathbf a}$ has been presented previously in \citet{Page2019}. 
Finally, he time $t$ in (\ref{eqn:u_guess}) is selected as $\hat{t} = \arg$ \hspace{-0.7mm} $\min_t |\mathbf u_g(t) - \mathbf u(t)|$. 

\section{Turbulent orbits}
Motivated by the success of the DMD approach at identifying properties of the edge-state UPO on short time windows, 
we now apply our method to turbulent data and attempt to converge UPO guesses generated as described above with a Newton solver.
To allow for a direct comparison to recurrent flow analysis, we perform many DMD computations on a short time window which is passed 
through a continuous turbulent trajectory of length $>1000$ advective time units.

\subsection{Recurrent flow analysis}
\begin{figure}
    \centering
    \includegraphics[width=0.97\textwidth]{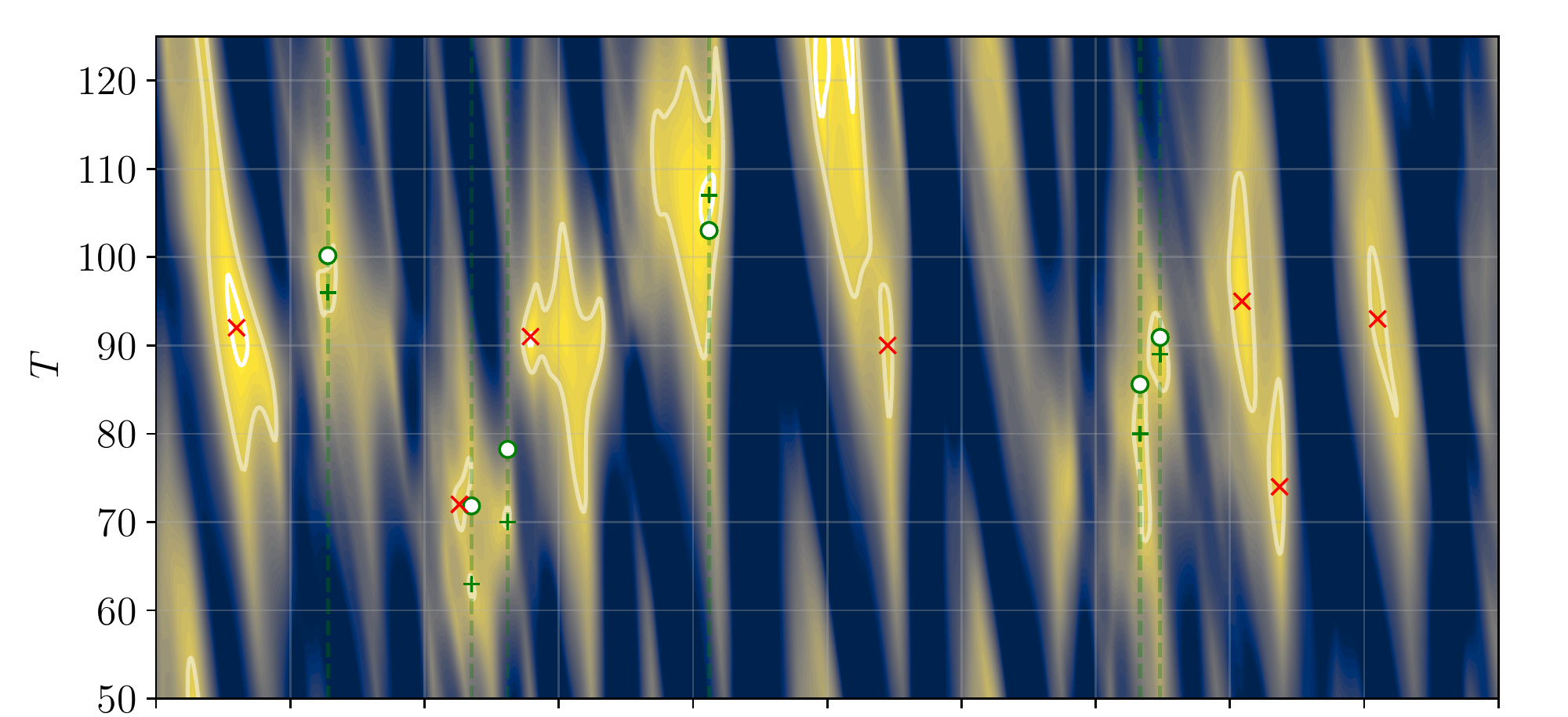}
    \includegraphics[width=0.97\textwidth]{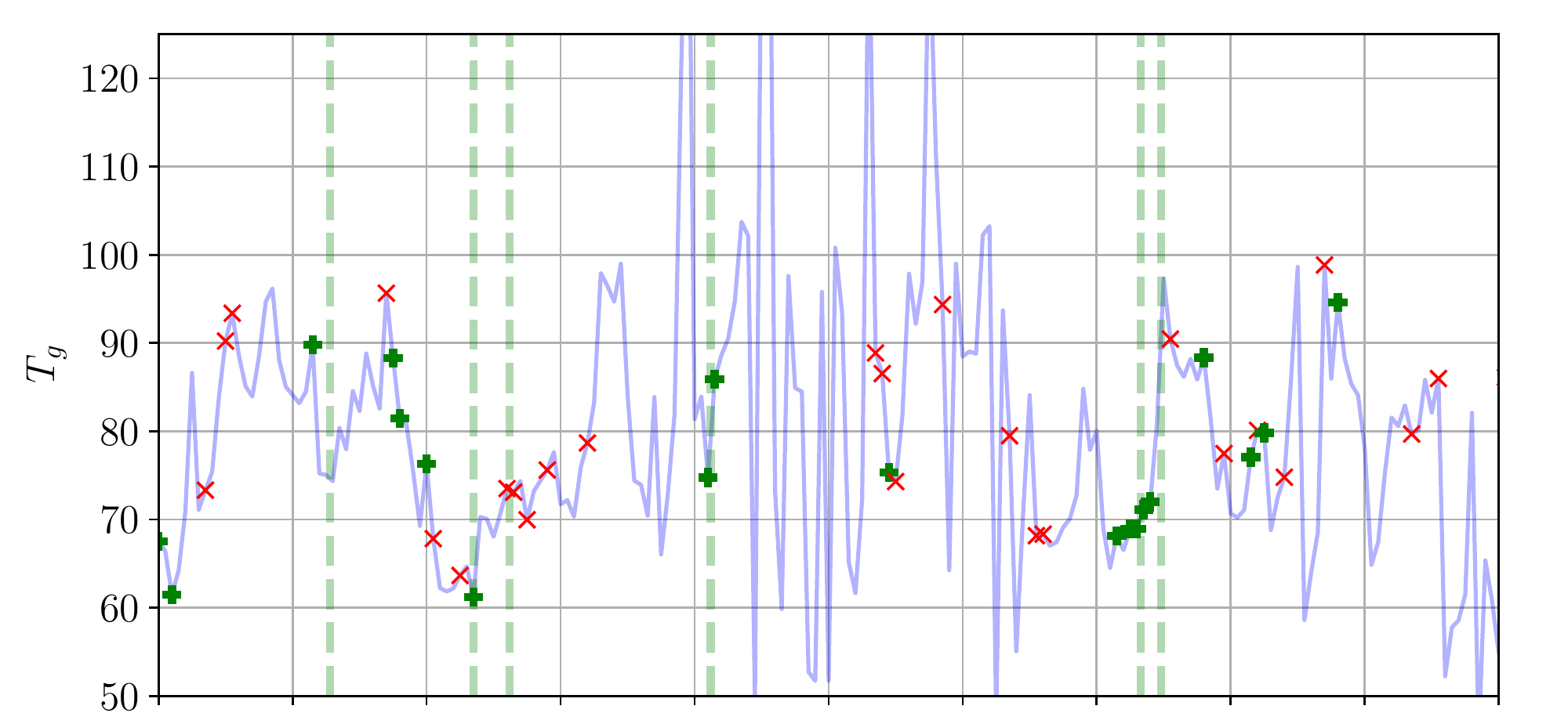}
    \includegraphics[width=0.972\textwidth]{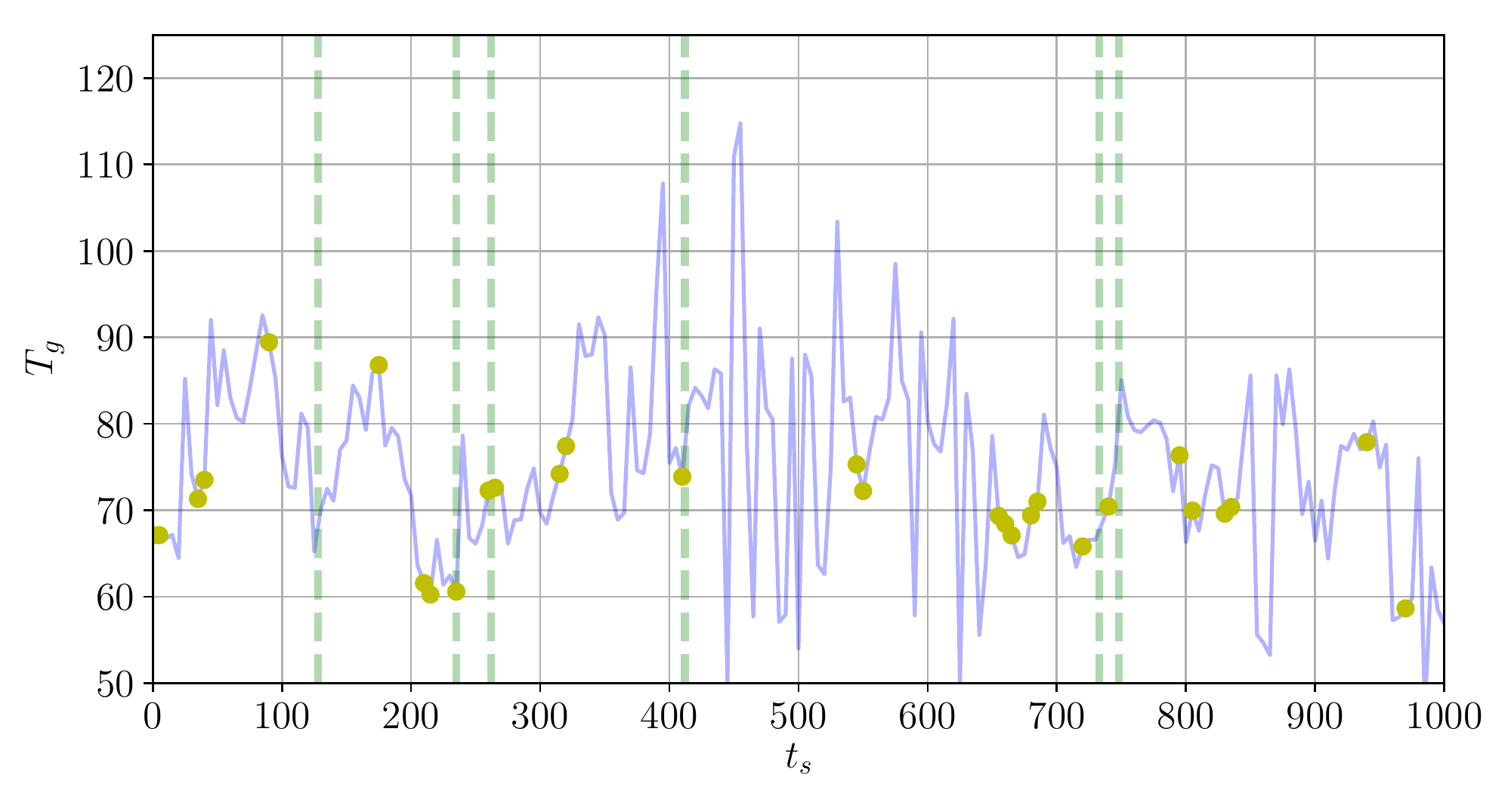}
    \caption{Caption on next page.}
    \label{fig:guesses_T60}
\end{figure}
\addtocounter{figure}{-1}
\begin{figure}
    \caption{(Figure on previous page).
    Summary of UPO guesses generated via recurrent flow analysis (top) and the DMD procedure outlined in the text performed on time
    windows of length $T_w=60$ using either 2 (centre) or 3 (bottom) harmonics to estimate a period $T_g$.  
    In the recurrence plot, green crosses identify successful guesses, while filled circles indicate the period of the converged solution.
    Red crosses are guesses that failed to converge.
    Note that the solid white contours identify values of $R=0.05$ and $R=0.07$. 
    DMD results are plotted against the start time of the observations, $t_s$, with a time difference $\Delta_w = 5$ separating subsequent calculations.
    $M=100$ snapshot pairs were used in each calculation, with $\delta t=1$.
    Points highlighted by symbols on the curves indicate where the value $\varepsilon_{\omega}$ dropped below a prescribed threshold ($\varepsilon=10^{-3}$ for 2 harmonics and $\varepsilon=5\times 10^{-3}$ for 3 harmonics). 
    All points satisfying this threshold in the two harmonic case were then fed into a Newton-Raphson algorithm: bold green crosses identify convergences, red crosses are failures.
    Yellow spots in the lowest plot indicate where the 3-harmonic procedure flagged likely UPOs (no attempt was made to converge these).
    }
\end{figure}

The results of a recurrent flow analysis are summarised in the top panel of figure \ref{fig:guesses_T60}, where we report contours 
of $R(t_s,T)$ (see equation \ref{eqn:recur}). 
Local minima which fall below a threshold value of $R_{\text{thresh}}=0.07$ are supplied as guesses in a Newton-Krylov algorithm.
Of the 13 guesses highlighted, 6 converged (green $+$, while red $\times$ indicate failures).

It is notable that the guessed period, determined from the $T$ at which the local minimum in $R(t_s,T)$ occurs, can often differ
significantly from the period of the converged solution (note the vertical distance between green $+$ and green $o$).
For example, the (good) guesses flagged at $t= 235$ and $t=262$ converged to periodic orbits with relative errors in the period against the guess of $12.3$\% 
and $10.5$\% respectively. 
As we shall see below, the combination of the state $\mathbf u(t)$ and period $T$ which correspond to a local minima in 
$R(t,T)$ can often result in failure in the Newton algorithm even if there is a nearby UPO. 
This is particularly evident in figure \ref{fig:guesses_T60} for $t > 800$, 
where recurrent flow analysis fails to identify and successful guesses,
while 
our new approach results 
in three converged UPOs.

\subsection{Dynamic mode decomposition}
\begin{table}
\setlength{\tabcolsep}{12pt}
  \begin{center}
  \begin{tabular}{lccccc}
      $t_s$  & $T_g$   &   $T_{\omega}$ & \% error in $T_g$ & \texttt{channelflow} $T_{\omega}$  & $T^{RFA}_{\omega}$($t_s$, \% error) \\[3pt]
         0   & 67.69   &   86.43        &   21.6            & -                                  & -                                   \\
        10   & 62.12   &   85.60        &   27.5            & 85.27                              & -                                   \\
        55   & 92.64   &   95.76        &    3.3            & -                                  & -                                   \\
       115   & 92.42   &   97.62        &    5.3            & -                                  & 100.17(128,4.2)                     \\
       175   & 84.35   &   77.25        &    9.1            & -                                  & -                                   \\
       180   & 81.47   &   76.54        &    6.4            & -                                  & -                                   \\
       200   & 75.04   &   78.23        &    4.1            & -                                  & -                                   \\
       235   & 61.38   &   71.83        &    14.5           & -                                  & 71.83(235,12.3)                     \\
       410   & 72.86   &   95.76        &    23.9           & -                                  & 103.01(412,3.7)                     \\
       415   & 84.19   &   104.28       &    23.9           & -                                  & 103.01(412,3.7)                     \\
       545   & 76.13   &   86.43        &    11.9           & -                                  & -                                   \\
       715   & 68.12   &   64.95        &    4.8            & -                                  & -                                   \\
       725   & 68.99   &   94.76        &   27.2            & -                                  & -                                   \\
       730   & 70.19   &   64.95        &    8.0            & -                                  & 85.60(733,6.5)                      \\
       735   & 71.31   &   88.61        &    17.3           & 87.89                              & 85.60(733,6.5)                      \\
       740   & 71.58   &   88.20        &    19.5           & 88.90                              & 85.60(733,6.5)                      \\
       780   & 88.73   &   90.71        &    3.3            & 90.31                              & -                                   \\
       815   & 76.19   &   78.23        &    2.6            & -                                  & -                                   \\
       825   & 79.71   &   70.12        &   13.7            & -                                  & -                                   \\
       880   & 95.14   &  104.28        &    8.8            & -                                  & -                                   \\
  \end{tabular}
      \caption{Periodic orbits found from the DMD results reported in figure \ref{fig:guesses_T60}.
      The guessed period of the UPO, $T_g$, was determined from $\omega_f(n=2)$ (equation \ref{eqn:freq_estimate}).
      In the penultimate column, the UPOs found from DMD-generated guesses 
      are connected to known UPOs converged in the slightly shorter $L_x=1.75 \pi$ box listed on \texttt{channelflow.org} 
      (see also the discussion at the start of \S2).
      The final column shows a comparison to recurrent flow analysis,  
      $T_{\omega}^{RFA}$ indicates the period of any UPOs found with that method for start times close to that used in the DMD.
      }
  \label{tab:DMD}
  \end{center}
\end{table}
The output of the DMD-based UPO detection algorithm outlined in \S\ref{sub:upo_id} is shown directly below the recurrence plot 
in figure \ref{fig:guesses_T60}, with guessed periods $T_g$ reported based on both two and three harmonics.
To generate these results, DMD computations were performed with a fixed observation window of length $T_w=60$, which was moved
through the time series in steps of $\Delta_w=5$.
Highlighted points along the curves $T_g(t_s)$ indicate that the value of $\varepsilon_{\omega}$ (equation \ref{eqn:spec_err})
dropped below a specified threshold.
These points tend to be arranged in clusters with similar guesses for the period $T_g$, 
consistent with the trajectory briefly shadowing a UPO before being flung 
out of a particular Koopman expansion zone \citep{Page2019}.
The detection threshold was set at $\varepsilon_{\omega}=10^{-3}$ for $n=2$ and $\varepsilon_{\omega}=5\times 10^{-3}$ for $n=3$;
the threshold was increased for $n=3$ since resolution of the third harmonic is more challenging.
The increase in the number of harmonics used in the search for UPOs results in a drop in the number of proposed guesses,
most notably for guesses with long periods. 
For example, the number of guesses with $T_g>80$ falls from 16 to to 3. 
However, in general there is good correspondence for the locations $\{t_s^i\}$ of the possible UPOs between the $n=2$ and $n=3$ results. 

The choice of observation window length $T_w=60$ was motivated by two factors:
(i) if the time window is too long and the turbulence visits the neighbourhoods of multiple UPOs, the spectrum from the DMD 
will likely be unrelated to any of them \citep[see the discussion in the introduction and][]{Page2019} and
(ii) $T_w=60$ is shorter than many of the known UPOs in this box, most of which have periods $T_{\omega}>75$ \citep{Cvitanovic2010}
and will test our method's ability to identify and converge solutions without a near recurrence. 
One immediate consequence of the choice $T_w=60$ is a loss of resolution of the longest UPO found by the recurrent flow analysis at $t_s=412$
with period $T_{\omega}=103.01$, though notably the DMD-based algorithm does flag a pair of likely 
UPOs with different periods close to this point -- a behaviour which is discussed in more detail below. 
However, in general DMD identifies guesses with similar periods to the UPOs spotted in the recurrent flow analysis.
Some heuristics around characteristics possessed by `good' DMD guesses are discussed in \S4, though ultimately we suspect the selection of $T_w$ 
in new applications is something that will have to be determined via trial-and-error subject to the tradeoff described above.

All of the guesses in figure \ref{fig:guesses_T60} identified by DMD in the 2-harmonic case 
were supplied as initial conditions in a Newton-Krylov routine.
Of the 46 guesses highlighted, 20 converged to periodic orbits.
This total includes UPOs found in regions where the recurrent flow analysis indicated the presence of a structure, 
but also in regions where no near-recurrent episodes were detected; note the guesses for $t_s\lesssim 50$ and for $150 \lesssim t_s \lesssim 200$.
The various UPOs converged from the DMD-based guesses identified in figure \ref{fig:guesses_T60} are 
summarised in Table \ref{tab:DMD},
and we will now discuss differences between these results and the recurrent flow analysis in some detail.
Our discussion will explore (i) cases where a UPO is found using recurrent flow analysis but the DMD-generated guess
fails to converge;
(ii) cases where the DMD predictions and recurrence predictions differ and different UPOs are converged;
(iii) cases where DMD and a recurrence analysis show similar predictions but only the DMD guess converges and finally
(iv) cases where DMD yields a converged UPO in the absence of a near recurrence.

(i) There are two instances where both the recurrent flow analysis and our DMD method indicate a likely UPO,
but only the guess from the recurrent flow analysis converges in the Newton solver ($t_s\approx 262$ and $t_s\approx 748$ in 
figure \ref{fig:guesses_T60}).
Intriguingly, the guesses for the period generated from the DMD are more accurate than those identified in the recurrence plot.
This suggests that the DMD-mode based guesses constructed using equation (\ref{eqn:u_guess}) constitute poor initial conditions
for the Newton solver.
In instances like these with both a near recurrence and a DMD indicator, a more robust approach could potentially combine 
elements of both methods, perhaps with a guess that is a concatenation of the state itself and the period
coming from DMD, i.e. $[\mathbf u(t_i), T_g]^T$.

(ii) On several occasions both the DMD and recurrent flow analysis identify possible UPOs at the same approximate $t_s$, 
but the converged solutions differ.
In some of these cases, it can be argued that one approach is `more accurate' than the other.
For example, compare the UPO with period $T_{\omega}=103.01$ found from a near recurrence at $t_s=412$ to the DMD predictions 
at $t_s\in \{410, 415\}$.
As described above, the short DMD time window is unable to accurately parameterise the long UPO with significant errors in the 
predicted period(s).
However, both DMD-based guesses do converge -- to UPOs with different periods that  are commensurate with the period obtained via recurrent flow analysis.
The change in period in the Newton iterations is striking and is not something we have observed in guesses from recurrent flow analysis.
There are other times where the UPOs found via DMD and recurrence differ (e.g. $t_s\approx 128$, $t_s\approx 730$) but the differences in
the predicted and true periods are less extreme and it is difficult to argue which guess was ``better''.
Here, there is a plausible connection with the project-then-search method described by \citet{Ahmed2017} for finding equilibria.
In that method, known exact solutions are projected onto a few resolvent modes before being input as new guesses to a Newton solver,
resulting in a large number of new equilibria which are qualitatively similar to the known solutions.
In our approach, we initialise the Newton solver with a guess constructed from just a few (5 for the 2 harmonic cases) DMD modes.

These observations raise an interesting ambiguity:
given that there appear to be groups of UPOs which are visually very similar (i.e. constitute a closed loop 
of the same physical processes) with very similar periods, many of which can be accessed from a Newton search with 
slightly different initial conditions, can we determine which solution the turbulent flow is actually shadowing?
Although we don't explore this here, more insight could potentially be gained by employing a longer observation window (e.g.
taken to be slightly longer than any identified near recurrence) and by using more harmonics to generate the initial guess.

There is less ambiguity in cases where either the recurrent flow analysis guess fails (iii), 
or where no recurrence is observed (iv). 
In these situations the DMD-based approach offers significant advantages. 
An example of the former scenario is seen in the DMD-based successes for $t_s\in \{815, 825\}$, where 
two near recurrences are found nearby but neither converges.
In both of these cases, the period generated by the DMD-based approach differs non-negligibly from 
the time $T$ at which $R(t_s,T)$ has a local minima. 
This is another scenario where the two methods can work well together to converge exact solutions where they 
may previously have been dismissed.


\begin{figure}
\centering
\includegraphics[width=0.529\textwidth]{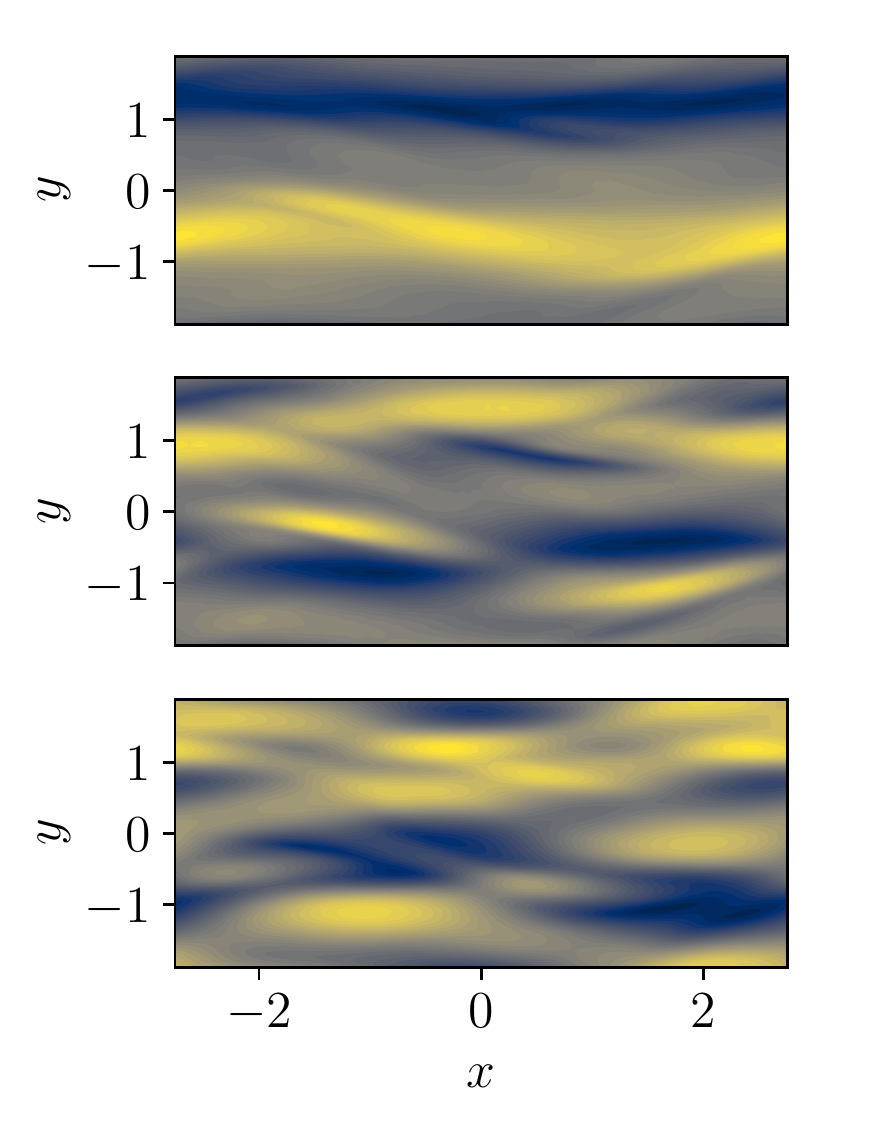}
\includegraphics[width=0.431\textwidth]{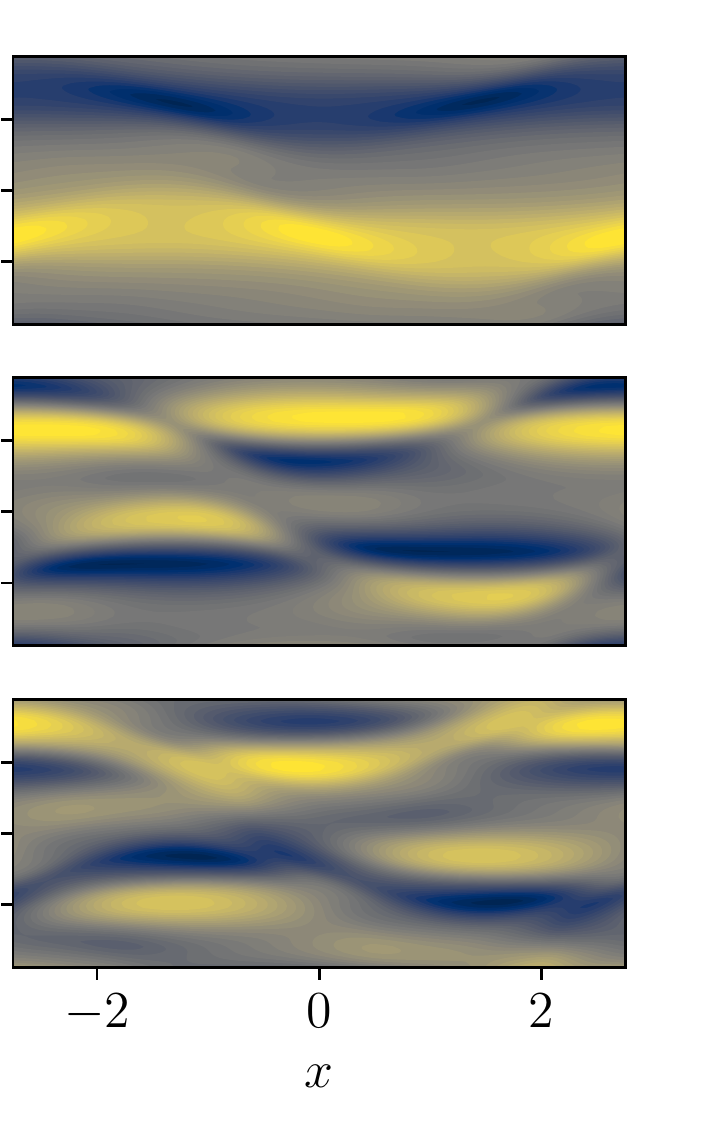}
\caption{Contours of (the real part of) the streamwise velocity for the first three DMD modes for (left) 
the time window $T_w=60$ starting at $t_s=780$ (see Table \ref{tab:DMD}) and (right) the converged UPO with period
$T_{\omega}=71.83$. In both cases, the number of snapshot pairs is $M=100$, with $\delta t = 1$.}
\label{fig:streaks}
\end{figure}
\begin{figure}
\centering
\includegraphics[width=0.529\textwidth]{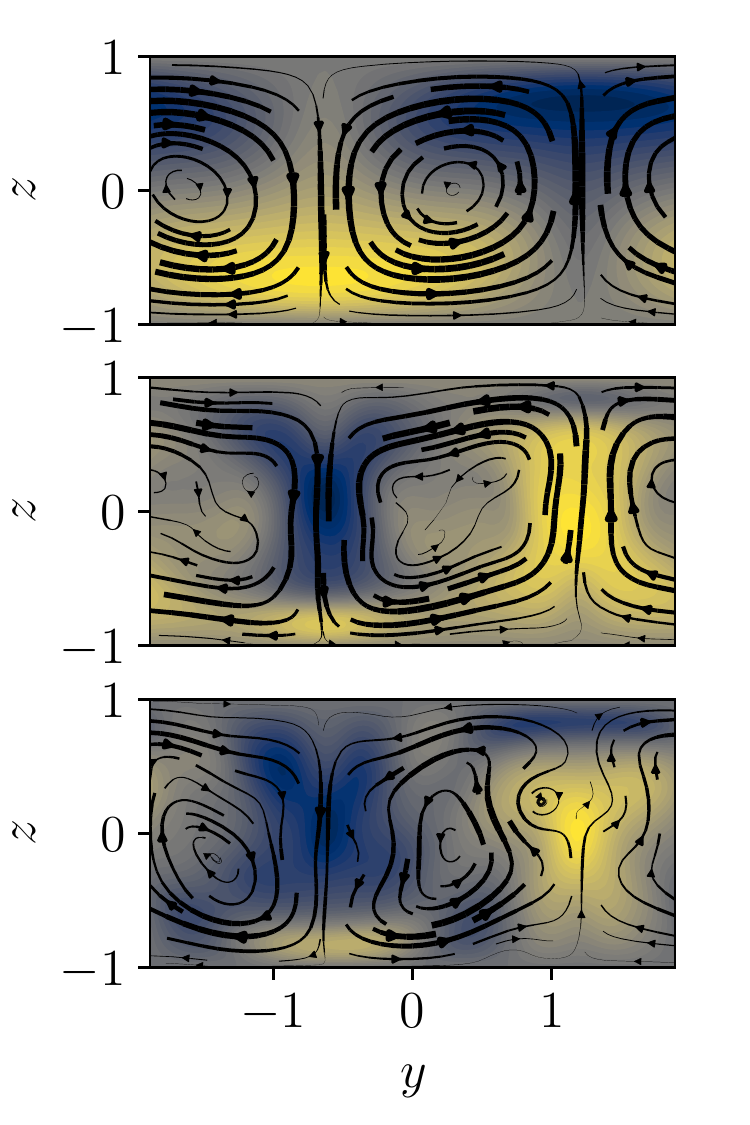}
\includegraphics[width=0.431\textwidth]{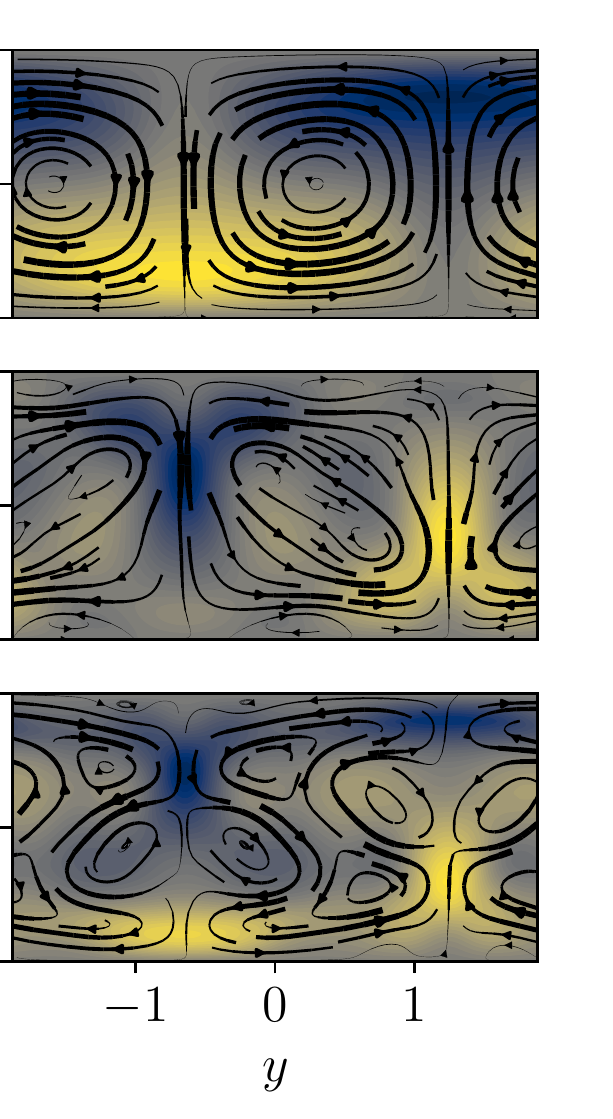}
\caption{As figure \ref{fig:streaks}, but now showing the streamwise-averaged streamwise velocity (contours)
and the streamwise average $y-z$ streamfunction (lines and arrows).}
\label{fig:rolls}
\end{figure}
Examples of the second scenario (iv), a converged UPO from DMD \emph{without a corresponding near recurrence}, are observed at $t_s\in\{180, 200, 780\}$.
These results are also notable due to the quality of the guess (e.g. the relative error in the period at $t_s=780$ is $3.3$\%).
The guess at $t_s=780$ is particularly interesting since the DMD window corresponds to only $2/3$ of the full period,
and in figures \ref{fig:streaks} and \ref{fig:rolls} we examine it further,
comparing the DMD modes used to build the UPO guess to the true Koopman modes of the UPO (the coefficients of
a Fourier series).
There is good qualitative agreement between the modes, providing further evidence 
that DMD of modest-duration turbulent trajectories can really see the essence of a nearby exact coherent structure.

It is worth emphasising that this final class of guesses could not be obtained another way as there is 
no near recurrence in the data, and so this method appears to offer some promise at higher Reynolds numbers
where the UPOs tend to be more unstable.
A common feature of the various guesses from DMD without a near recurrence is that only the first two harmonics
are resolved accurately (see the difference in the bottom two panels of \ref{fig:guesses_T60}).
This behaviour was also observed for the simple edge state examined in \S\ref{sec:edge},
where we observed improved resolution of the higher-frequency Koopman modes as the length of the observation window was increased.
However, for turbulent trajectories the observation window cannot be increased arbitrarily and would require 
a careful adjustment to avoid the inclusion of snapshots where the flow has left the neighbourhood around the UPO 
where the local Koopman decomposition holds.

\section{Discussion}
In this section we briefly examine some features of our initial guesses to provide some insight into what constitutes a good guess for UPO.
We do this by exploring how well the short-time DMD computations can see Koopman eigenfunctions related to the structure of interest
and by performing DMD on the converged turbulent UPOs themselves to identify a rough rule-of-thumb on where the best DMD 
time windows tend to be located in state space.

\begin{figure}
    \centering
    \includegraphics[width=0.95\textwidth]{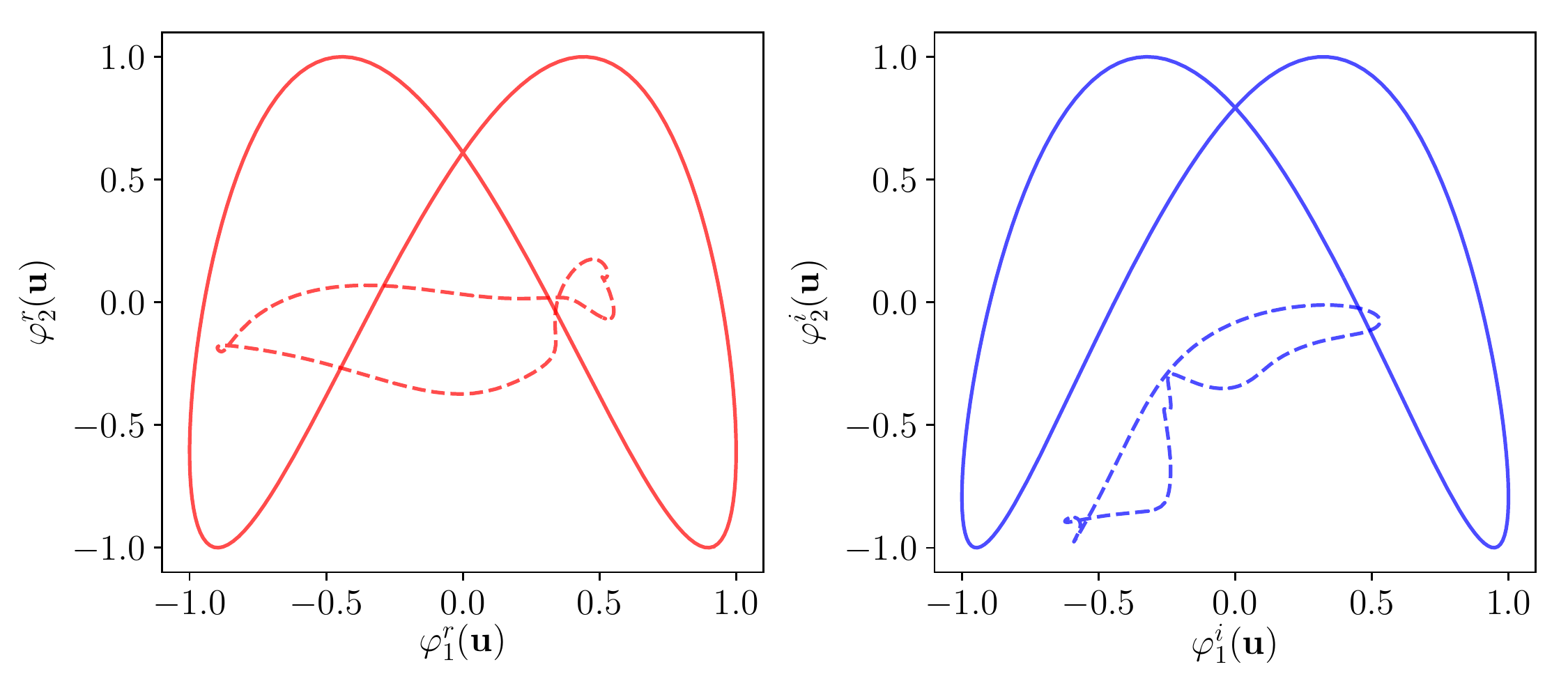}
    \includegraphics[width=0.95\textwidth]{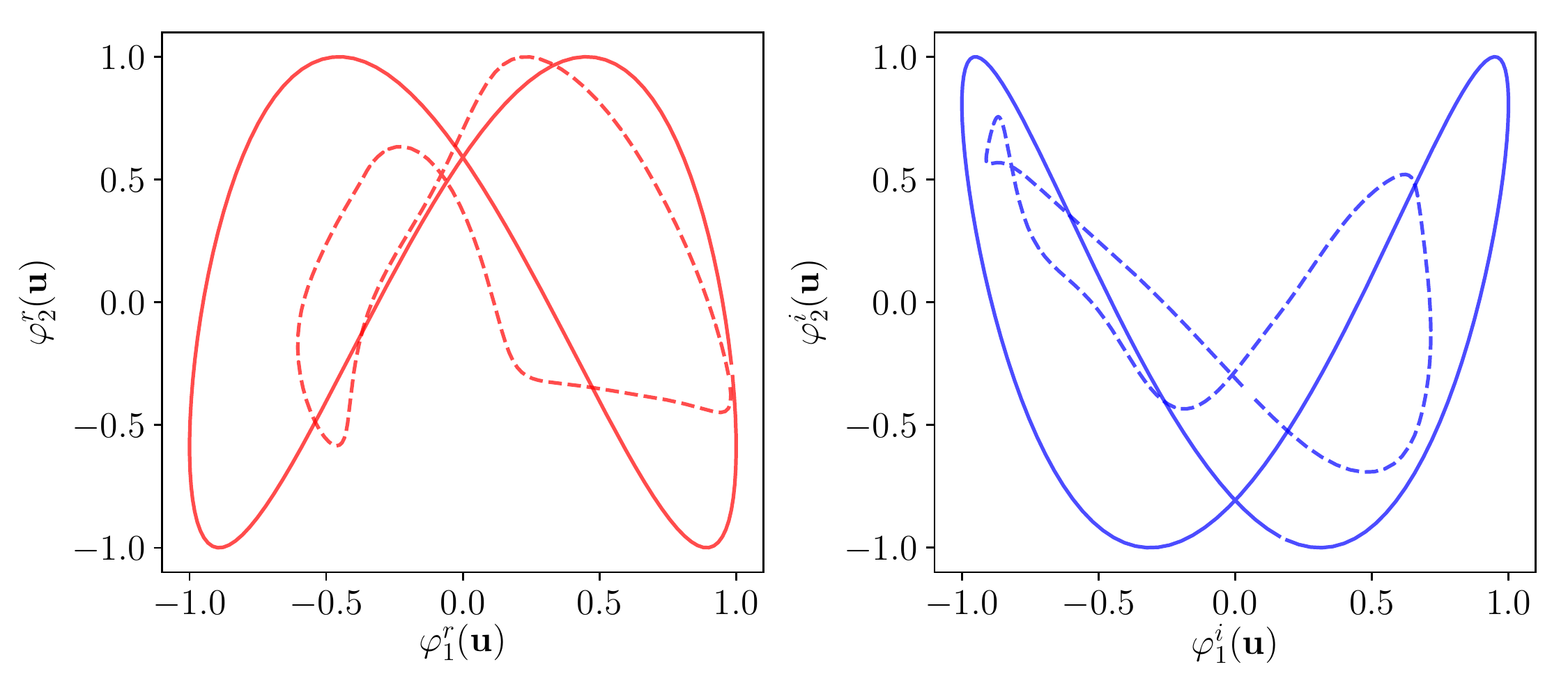}
    \caption{Real and imaginary components of approximate Koopman eigenfunctions, $\varphi = \varphi^r + \zi \varphi^i$, 
    corresponding to the eigenvalues $\lambda_1 \approx \zi\omega_f$ and $\lambda_2\approx 2\zi\omega_f$,
    obtained in DMD applied to a converged UPO (solid lines) and 
    the fixed turbulent window $T_w=60$ that converged to that same UPO (dashed lines).
    In both cases the eigenfunction is evaluated along the UPO itself.
    (Top) The initial guess at $t_s=10$ (see table \ref{tab:DMD}) and the converged UPO $T_{\omega}=85.60$.
    (Bottom) The initial guess at $t_s=235$ and the converged UPO $T_{\omega}=71.83$.
    }
    \label{fig:efns}
\end{figure}
The idea that DMD can identify the signature of UPOs in a long trajectory is based on the fact
that these exact coherent structures possess a local Koopman decomposition which is related to both
the period of the solution itself and the eigenvalues governing the dynamics within its linear subspace
\citep{Mezic2017}.
For short time windows, DMD can typically only resolve a few eigenvalues related to the structure itself (the UPO's
Fourier harmonics) as seen in \S 3 \citep[see also][]{Page2019}.
The `quality' of a particular DMD-based guess, and some indication as to whether the convergence of a UPO occurred due 
to luck, can be examined by computing approximations to the Koopman eigenfunctions (see equation \ref{eqn:efns})
using left eigenvectors from the short-time-window DMD,
\begin{equation}
    \varphi^{DMD}_{\lambda_j}(\mathbf u) = (\mathbf w_j^{DMD})^H \boldsymbol \psi(\mathbf u),
\end{equation}
and comparing them to those obtained from the converged UPO itself,
\begin{equation}
    \varphi_{\lambda_j}(\mathbf u) = (\mathbf w_j^{UPO})^H \boldsymbol \psi(\mathbf u).
\end{equation}

The above procedure is performed for two periodic orbits and their original guesses in figure \ref{fig:efns}, where the velocity field used in the
projections is that of the UPO itself.
One of the guesses examined was selected due to its particularly poor esimate of the period of the UPO (at $t_s=10$, $T_g=62.12$ and $T_{\omega} = 85.60$, a relative
error of $27.5$\%) and the resolution of the Koopman eigenfunctions is also poor.
In particular, the projection shows sharp gradients and appears almost discontinuous in places. 
It does not at all resemble the true Koopman eigenfunction extracted from the UPO itself.
On the other hand, the second guess was considerably more accurate and the approximation to the Koopman eigenfunctions is 
qualitatively much better.
For other UPOs we have observed similar behaviour (not shown); 
that more accurate initial guesses (as measured by the predicted period) tend to produce 
Koopman eigenfunctions which resemble qualitatively those of the UPO itself.

\begin{figure}
    \centering
    \includegraphics[width=0.95\textwidth]{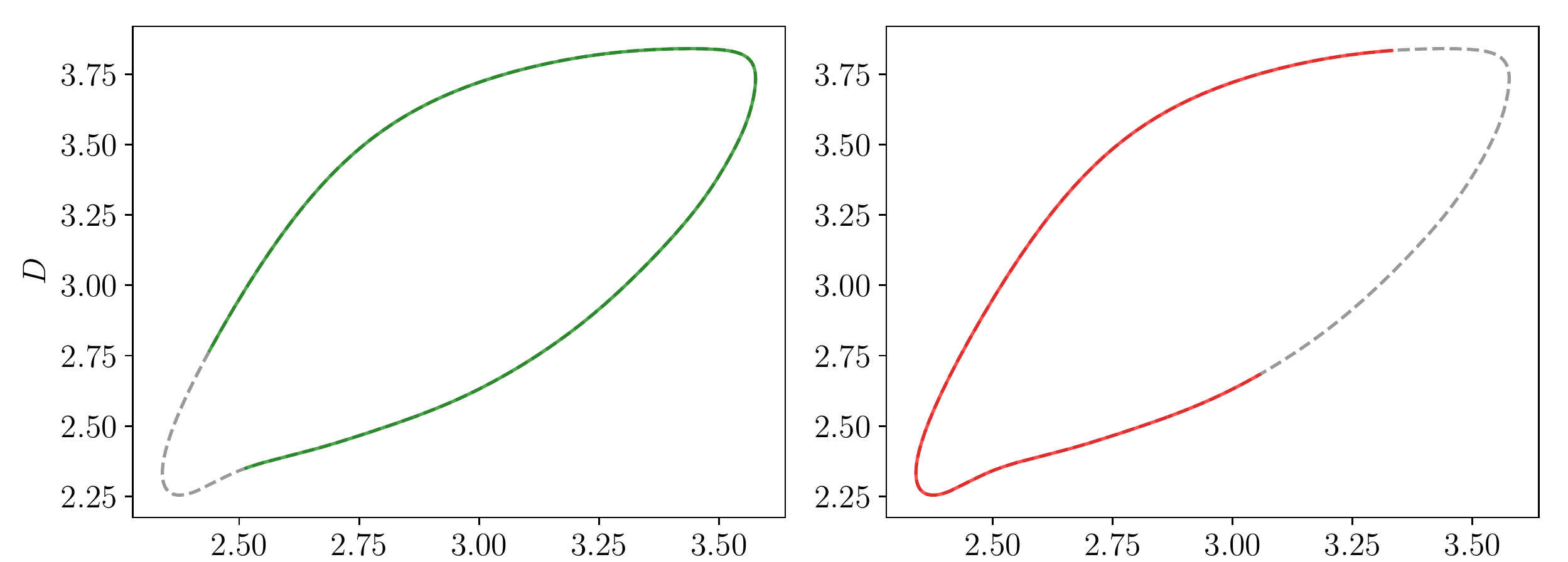}
    \includegraphics[width=0.95\textwidth]{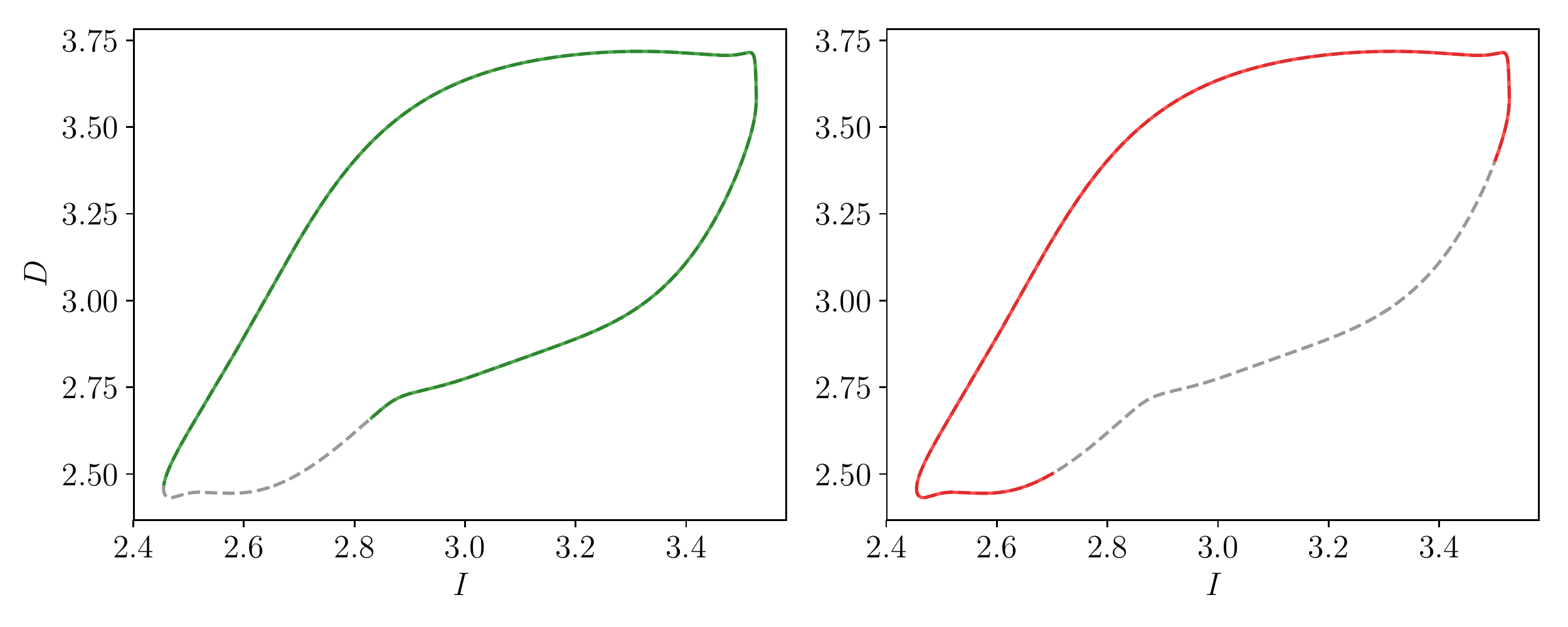}
    \caption{
        ``Best'' (left, in green) and ``worst'' (right, in red) locations for a DMD time window of length $T_w=60$ as measured by the 
        difference in predicted period from that of the true UPO, $|T_g - T_{\omega}|/T_{\omega}$, visualised in  
        a two-dimensional projection of energy production against dissipation (both normalized by their values in
        laminar flow).
        The computations are performed on the UPOs themselves, and the full UPO is shown in dashed grey. 
        The predicted period $T_g=2\pi/\omega_f$, where $\omega_f$ was computed from the first two harmonics.
        (Top) background UPO has period $T_{\omega}=85.60$.
        (Bottom) background UPO has period $T_{\omega}=94.76$.
    }
    \label{fig:ID_measure}
\end{figure}
Computation of the Koopman eigenfunctions along the UPO itself is only possible after the solution has been converged,
and it is natural to speculate whether there are additional indicators of guess quality that can be assessed before
the Newton solver is initialised.
While the eigenspectrum-based measure $\varepsilon_{\omega}$ (equation \ref{eqn:spec_err}) provides a relatively 
robust classifier to identify sensible initial guesses, we have observed some additional heuristics that can be used to screen 
a large number of initial guesses.
When examinining a long time series with short DMD windows, 
perhaps the most obvious clue is the observation of `clusters' of similar guesses (in terms of period $T_g$) in time.
This is evident in some of the successful guesses highlighted in figure \ref{fig:guesses_T60} and could 
be further emphasised by shrinking the timestep between subsequent DMD calculations (in figure \ref{fig:guesses_T60} this is $\Delta_w=5$). 

Another interesting feature of good guesses from short DMD windows is that they 
tend to share certain characteristics in terms of their location 
in an energy production-dissipation plane.
To demonstrate this effect, we perform many DMD calculations with fixed window length $T_w=60$ on two of the longer converged UPOs
and plot the `best' and `worst' time windows in figure \ref{fig:ID_measure}, where the criteria used to rank the quality of the DMD
spectra was the error in the guessed period (using the first two harmonics), $|2\pi/\omega_f(n=2) - T_{\omega}|/T_{\omega}$.
The best guesses tend to sample the upper-right quadrant (the fast, high production, high dissipation region) at the expense 
of the slower, gentler region in the lower-left corner.
Note that for longer orbits, good estimates (relative error $\lesssim 5$\%) can also be obtained on trajectories 
which mostly shadow the portion of the curve $D<I$ (not shown) -- 
the key requirement appears to be the inclusion of the most extreme values of production and dissipation.

We note that the presentation in this manuscript has focused on one particular time window length and DMD design
in order to introduce the method in a clear and systematic way.
There is clearly scope for the results to be improved by adjusting the DMD design based on local performance.
For example, where near recurrences are observed it would make sense to increase the time window to the predicted period,
while for cases where no near recurrence is observed but DMD does predict a UPO, it would be interesting to search for a `best' local guess
by small adjustments in window length and position.

\section{Conclusion}
We have presented a new method based on DMD to both identify the signature of nearby UPOs in time-series of turbulent flows, and to generate
robust initial guesses that can serve as inputs to a Newton-Krylov algorithm.
The approach is designed to be applied to short turbulent trajectories, with the existence of a nearby UPO deemed
`likely' when the DMD eigenvalue spectrum exhibits repeated harmonics of a fundamental frequency.
Initial conditions for the Newton solver are then built from these near-periodic DMD modes 
and a guess for the period based on the estimated fundamental frequency.

The ability of the method to function without the need for a near recurrence within the observation window was 
demonstrated on a simple UPO before we applied it to a long turbulent trajectory. 
With a modest time window (shorter than most known UPOs), 
the new method identified many more UPOs than were found with a recurrent flow analysis of the same dataset.
These solutions were found in regions where (i) a near recurrence was found and converged, (ii)  a near recurrence was identified but the Newton solver
failed to converge and (iii) where no near recurrence was flagged. 
These outcomes indicate that the method can be both a useful complementary tool 
and a potentially powerful alternative to recurrent flow analysis.
In particular, the success of the method in regions where no near recurrence occurs suggests it may be of use
at higher Reynolds numbers where the probability of shadowing a UPO for a full cycle drops.

An interesting behaviour was found at times where both the DMD based method and recurrent flow analysis flagged the probable existence of a UPO. 
In some cases, the DMD approach converged to a different solution than the guess from a recurrent flow analysis.
Often, the error in the predicted period for these alternate solutions was comparable to the recurrent flow analysis,
and given the sensitivity of the Newton method to initial conditions it is difficult to assert which one was indeed the 
solution being shadowed in phase space.
Motivated by the variable discrepancy between the predicted period and that of the converged solution, we also presented some 
analysis of the features of both `good' and `bad' initial guesses generated by the method.
Good guesses tend to require the DMD to see the most extreme dissipation events along an orbit at the expense 
of the slower, more gentle dynamics.

There are a number of interesting avenues open to further investigation.
The most natural question is whether the method can be extended to much higher Reynolds numbers to converge UPOs,
where recurrent flow analysis struggles to identify any guesses at all \citep[e.g. see][]{Chandler2013, Lucas2015}.
There are also some intriguing properties of the DMD-based approach that merit further study.
For example, DMD on short time windows performs ``best'' (i.e. the predicted period matches that of the underlying simple invariant set)
when it sees the faster, high-dissipation region of the UPO. 
In instances where the observation window does not have this property, can the performance of the DMD be improved, for instance
by modifying how snapshot pairs are distributed in time?

\section*{Acknowledgements}
We are very grateful to Yves Dubief for sharing his viscoelastic code with us.

\bibliographystyle{jfm}
\bibliography{upo_dmd}

\begin{thebibliography}{36}
\expandafter\ifx\csname natexlab\endcsname\relax\def\natexlab#1{#1}\fi
\def\au#1{#1} \def\ed#1{#1} \def\yr#1{#1}\def\at#1{#1}\def\jt#1{\textit{#1}}
  \def\bt#1{#1}\def\bvol#1{\textbf{#1}} \def\vol#1{#1} \def\pg#1{#1}
  \def\publ#1{#1}\def\arxiv#1{#1}\def\org#1{#1}\def\st#1{\textit{#1}}

\bibitem[Ahmed \& Sharma(2017)]{Ahmed2017}
{\sc \au{Ahmed, M.~A.} \& \au{Sharma, A.~S.}} \yr{2017}  \at{{New equilibrium
  solution branches of plane Couette flow discovered using a
  project-then-search method}}.  \jt{arXiv 1706.05312} .

\bibitem[Arbabi \& Mezi\'c(2017)]{Arbabi2017}
{\sc \au{Arbabi, H.} \& \au{Mezi\'c, I.}} \yr{2017}  \at{{Study of dynamics in
  post-transient flows using Koopman mode decomposition}}.  \jt{Phys. Rev.
  Fluids}  \bvol{2},  \pg{124402}.

\bibitem[Bagheri(2013)]{Bagheri2013}
{\sc \au{Bagheri, S.}} \yr{2013}  \at{{Koopman-mode decomposition of the
  cylinder wake}}.  \jt{J. Fluid Mech.}  \bvol{726},  \pg{596--623}.

\bibitem[Brunton {\em et~al.\/}(2016)Brunton, Brunton, Proctor \&
  Kutz]{Brunton2016}
{\sc \au{Brunton, S.~L.}, \au{Brunton, B.~W.}, \au{Proctor, J.~L.} \& \au{Kutz,
  J.~N.}} \yr{2016}  \at{{Koopman invariant subspaces and finite linear
  repesentations of nonlinear dynamical systems for control}}.  \jt{PLoS ONE}
  \bvol{11}~(2).

\bibitem[Chandler \& Kerswell(2013)]{Chandler2013}
{\sc \au{Chandler, G.~J.} \& \au{Kerswell, R.~R.}} \yr{2013}  \at{Invariant
  recurrent solutions embedded in a turbulent two-dimensional {K}olmogorov
  flow}.  \jt{Journal of Fluid Mechanics}  \bvol{722},  \pg{554–595}.

\bibitem[Cvitanovi{\'c} {\em et~al.\/}(2016)Cvitanovi{\'c}, Artuso, Mainieri,
  Tanner \& Vattay]{ChaosBook}
{\sc \au{Cvitanovi{\'c}, P.}, \au{Artuso, R.}, \au{Mainieri, R.}, \au{Tanner,
  G.} \& \au{Vattay, G.}} \yr{2016} {\em Chaos: Classical and Quantum\/}.
  \publ{Copenhagen: Niels Bohr Inst.}

\bibitem[Cvitanovic \& Gibson(2010)]{Cvitanovic2010}
{\sc \au{Cvitanovic, P.} \& \au{Gibson, J.~F.}} \yr{2010}  \at{{ Geometry of
  the turbulence in wall-bounded shear flows: periodic orbits }}.  \jt{Physica
  Scripta}  \bvol{T142},  \pg{014007}.

\bibitem[Dubief {\em et~al.\/}(2005)Dubief, Terrapon, White, Shaqfeh, Moin \&
  Lele]{Dubief2005}
{\sc \au{Dubief, Y.}, \au{Terrapon, V.~E.}, \au{White, C.~M.}, \au{Shaqfeh, E.
  S.~G.}, \au{Moin, P.} \& \au{Lele, S.~K.}} \yr{2005}  \at{{New answers on the
  interaction between polymers and vortices in turbulent flows}}.  \jt{Flow,
  Turbulence and Combustion}  \bvol{74}~(4),  \pg{311--329}.

\bibitem[Hall \& Sherwin(2010)]{Hall2010}
{\sc \au{Hall, P.} \& \au{Sherwin, S.}} \yr{2010}  \at{Streamwise vortices in
  shear flows: harbingers of transition and the skeleton of coherent
  structures}.  \jt{Journal of Fluid Mechanics}  \bvol{661},  \pg{178–205}.

\bibitem[Hamilton {\em et~al.\/}(1995)Hamilton, Kim \& Waleffe]{Hamilton1995}
{\sc \au{Hamilton, J.~M.}, \au{Kim, J.} \& \au{Waleffe, F.}} \yr{1995}
  \at{Regeneration mechanisms of near-wall turbulence structures}.  \jt{Journal
  of Fluid Mechanics}  \bvol{287},  \pg{317–348}.

\bibitem[Jovanovi\'c {\em et~al.\/}(2014)Jovanovi\'c, Schmid \&
  Nichols]{Jovanovic2014}
{\sc \au{Jovanovi\'c, M.~R.}, \au{Schmid, P.~J.} \& \au{Nichols, J.~W.}}
  \yr{2014}  \at{{Sparsity-promoting dynamic mode decomposition}}.  \jt{Phys.
  Fluids}  \bvol{26},  \pg{024103}.

\bibitem[Kawahara(2005)]{Kawahara2005}
{\sc \au{Kawahara, G.}} \yr{2005}  \at{Laminarization of minimal plane couette
  flow: {g}oing beyond the basin of attraction of turbulence}.  \jt{Physics of
  Fluids}  \bvol{17},  \pg{041702}.

\bibitem[Kawahara \& Kida(2001)]{Kawahara2001}
{\sc \au{Kawahara, G.} \& \au{Kida, S.}} \yr{2001}  \at{Periodic motion
  embedded in plane couette turbulence: regeneration cycle and burst}.
  \jt{Journal of Fluid Mechanics}  \bvol{449},  \pg{291–300}.

\bibitem[Kawahara {\em et~al.\/}(2012)Kawahara, Uhlmann \& van
  Veen]{Kawahara2012}
{\sc \au{Kawahara, G.}, \au{Uhlmann, M.} \& \au{van Veen, L.}} \yr{2012}
  \at{The significance of simple invariant solutions in turbulent flows}.
  \jt{Annual Review of Fluid Mechanics}  \bvol{44}~(1),  \pg{203--225}.

\bibitem[Kerswell(2005)]{Kerswell2005}
{\sc \au{Kerswell, R.~R.}} \yr{2005}  \at{{ Recent progress in understanding
  the transition to turbulence in a pipe}}.  \jt{Nonlinearity}  \bvol{18},
  \pg{R17--R44}.

\bibitem[Koopman(1931)]{Koopman1931}
{\sc \au{Koopman, B.~O.}} \yr{1931}  \at{{Hamiltonian Systems and
  Transformations in Hilbert Space}}.  \jt{Proc. Nat. Acad. Sci.}
  \bvol{17}~(5),  \pg{315--318}.

\bibitem[Kutz {\em et~al.\/}(2016)Kutz, Brunton, Brunton \& Proctor]{DMDkutz}
{\sc \au{Kutz, J.~N.}, \au{Brunton, S.~L.}, \au{Brunton, B.~W.} \& \au{Proctor,
  J.~L.}} \yr{2016} {\em {Dynamic Mode Decomposition: Data-Driven Modeling of
  Complex Systems}\/}, 1st edn.  \publ{SIAM}.

\bibitem[Lucas {\em et~al.\/}(2017)Lucas, Caulfield \& Kerswell]{Lucas2017}
{\sc \au{Lucas, D.}, \au{Caulfield, C.~P.} \& \au{Kerswell, R.~R.}} \yr{2017}
  \at{Layer formation in horizontally forced stratified turbulence: connecting
  exact coherent structures to linear instabilities}.  \jt{Journal of Fluid
  Mechanics}  \bvol{832},  \pg{409–437}.

\bibitem[Lucas \& Kerswell(2015)]{Lucas2015}
{\sc \au{Lucas, D.} \& \au{Kerswell, R.~R.}} \yr{2015}  \at{Recurrent flow
  analysis in spatiotemporally chaotic 2-dimensional {K}olmogorov flow}.
  \jt{Physics of Fluids}  \bvol{27},  \pg{045106}.

\bibitem[Mezi\'c(2005)]{Mezic2005}
{\sc \au{Mezi\'c, I.}} \yr{2005}  \at{{Spectral Properties of Dynamical
  Systems, Model Reduction and Decompositions}}.  \jt{Nonlinear Dynam.}
  \bvol{41},  \pg{309--325}.

\bibitem[Mezi\'c(2013)]{Mezic2013}
{\sc \au{Mezi\'c, I.}} \yr{2013}  \at{{Analysis of Fluid Flows via Spectral
  Properties of the Koopman Operator}}.  \jt{Ann. Rev. Fluid Mech.}  \bvol{45},
   \pg{357--378}.

\bibitem[Mezic(2017)]{Mezic2017}
{\sc \au{Mezic, I.}} \yr{2017}  \at{{Koopman Operator Spectrum and Data
  Analysis}}.  \jt{arXiv 1702.07597} .

\bibitem[Mezic \& Banaszuk(2004)]{Mezic2004}
{\sc \au{Mezic, I.} \& \au{Banaszuk, A.}} \yr{2004}  \at{Comparison of systems
  with complex behavior}.  \jt{Physica D}  \bvol{197},  \pg{101--133}.

\bibitem[Page \& Kerswell(2018)]{Page2018}
{\sc \au{Page, J.} \& \au{Kerswell, R.~R.}} \yr{2018}  \at{{Koopman analysis of
  Burgers equation}}.  \jt{Physical Review Fluids}  \bvol{3},  \pg{071901(R)}.

\bibitem[{Page} \& {Kerswell}(2018)]{Page2019}
{\sc \au{{Page}, J.} \& \au{{Kerswell}, R.~R.}} \yr{2018}  \at{{Koopman mode
  expansions between simple invariant solutions}}.  \jt{arXiv 1811.05907} .

\bibitem[Rowley \& Dawson(2017)]{Rowley2017}
{\sc \au{Rowley, C.~W.} \& \au{Dawson, S. T.~M.}} \yr{2017}  \at{{Model
  Reduction for Flow Analysis and Control}}.  \jt{Ann. Rev. Fluid Mech.}
  \bvol{49},  \pg{387--417}.

\bibitem[Rowley {\em et~al.\/}(2009)Rowley, Mezi\'c, Bagheri, Schlatter \&
  Henningson]{Rowley2009}
{\sc \au{Rowley, C.~W.}, \au{Mezi\'c, I.}, \au{Bagheri, S.}, \au{Schlatter, P.}
  \& \au{Henningson, D.~S.}} \yr{2009}  \at{{Spectral analysis of nonlinear
  flows}}.  \jt{J. Fluid Mech.}  \bvol{641},  \pg{115--127}.

\bibitem[Schmid(2010)]{Schmid2010}
{\sc \au{Schmid, P.~J.}} \yr{2010}  \at{{Dynamic mode decomposition of
  numerical and experimental data}}.  \jt{J. Fluid Mech.}  \bvol{656},
  \pg{5--28}.

\bibitem[Schmid {\em et~al.\/}(2011)Schmid, Li, Juniper \& Pust]{Schmid2011}
{\sc \au{Schmid, P.~J.}, \au{Li, L.}, \au{Juniper, M.~P.} \& \au{Pust, O.}}
  \yr{2011}  \at{{Applications of the dynamic mode decomposition}}.
  \jt{Theoretical and Computational Fluid Dynamics}  \bvol{25},  \pg{249--259}.

\bibitem[Schneider {\em et~al.\/}(2007)Schneider, Eckhardt \&
  Yorke]{Schneider2007}
{\sc \au{Schneider, T.~M.}, \au{Eckhardt, B.} \& \au{Yorke, J.~A.}} \yr{2007}
  \at{Turbulence transition and the edge of chaos in pipe flow}.  \jt{Physical
  Review Letters}  \bvol{99},  \pg{034502}.

\bibitem[Tu {\em et~al.\/}(2014)Tu, Rowley, Luchtenburg, Brunton \&
  Kutz]{Tu2014}
{\sc \au{Tu, J.~H}, \au{Rowley, C.~W.}, \au{Luchtenburg, D.~M.}, \au{Brunton,
  S.~L.} \& \au{Kutz, J.~N.}} \yr{2014}  \at{{On dynamic mode decomposition:
  theory and applications}}.  \jt{J. Comput. Dynam.}  \bvol{1}~(2),
  \pg{391--421}.

\bibitem[Viswanath(2007)]{Viswanath2007}
{\sc \au{Viswanath, D.}} \yr{2007}  \at{Recurrent motions within plane
  {C}ouette turbulence}.  \jt{Journal of Fluid Mechanics}  \bvol{580},
  \pg{339–358}.

\bibitem[Waleffe(1997)]{Waleffe1997}
{\sc \au{Waleffe, F.}} \yr{1997}  \at{{ On a self-sustaining process in shear
  flows}}.  \jt{Physics of Fluids}  \bvol{9},  \pg{883--900}.

\bibitem[Wang {\em et~al.\/}(2007)Wang, Gibson \& Waleffe]{Wang2007}
{\sc \au{Wang, J.}, \au{Gibson, J.} \& \au{Waleffe, F.}} \yr{2007}  \at{Lower
  branch coherent states in shear flows: {t}ransition and control}.
  \jt{Physical Review Letters}  \bvol{98},  \pg{204501}.

\bibitem[Williams {\em et~al.\/}(2015)Williams, Kevrekidis \&
  Rowley]{Williams2015}
{\sc \au{Williams, M.~O.}, \au{Kevrekidis, I.~G.} \& \au{Rowley, C.~W.}}
  \yr{2015}  \at{{A Data-Driven Approximation of the Koopman Operator:
  Extending Dynamic Mode Decomposition}}.  \jt{J. Nonlinear Sci.}
  \bvol{25}~(6),  \pg{1307--1346}.

\bibitem[Willis {\em et~al.\/}(2013)Willis, Cvitanovic \& Avila]{Willis2013}
{\sc \au{Willis, A.~P.}, \au{Cvitanovic, P.} \& \au{Avila, M.}} \yr{2013}
  \at{{Revealing the state space of turbulent pipe flow by symmetry
  reduction}}.  \jt{Journal of Fluid Mechanics}  \bvol{721},  \pg{514--540}.

\end{thebibliography}
\end{document}